\documentclass{arxivversion}
\usepackage{graphicx} 
\usepackage{pgfplots}
\usepackage[latin9]{inputenc}
\usepackage{verbatim}
\usepackage{babel}
\usepackage{xcolor}
\usepackage{soul}
\usepackage{amsmath}
\usepackage{tikz}
\usepackage{float}
\pgfplotsset{compat=1.18}
\usepackage{bm}
\usepackage{csquotes}
\usepackage[makeroom]{cancel}
\usepackage[dvipsnames]{xcolor}
\usepackage{amssymb}
\usepackage{mathrsfs}
\usepackage{dcolumn}

\begin{document}

\articletype{Paper} 

\title{Nonlinear evolution of perturbatively driven kinetic instabilities far from marginal stability}

\author{Emma G. Devin$^{1,2}$ and Vin\'{i}cius N. Duarte$^{1,2}$}

\affil{$^1$Princeton Plasma Physics Laboratory, Princeton, USA}

\affil{$^2$Department of Astrophysical Sciences, Princeton University, Princeton, USA}

\email{edevin@pppl.gov, vduarte@pppl.gov} 


\begin{abstract}
    Kinetic instabilities develop when a system's distribution function deviates from thermal equilibrium in such a way that allows free energy from that distribution to drive resonant modes. These instabilities occur in many systems, such as fusion and astrophysical plasmas, neutral fluids, and self-gravitating systems. Motivated by the case of Alfv\'enic instabilities driven by minority populations of energetic particles in tokamak plasmas, we consider a kinetic instability in the presence of sources and sinks and with perturbative drive far from its instability threshold, where a theoretical description of the time evolution has not yet been established. In cases with steady-state saturation levels, we find that the mode first evolves linearly, driven by the positive distribution gradient, then undergoes a fast, strongly nonlinear transition where the distribution function slope is completely flattened around the resonance. The system then evolves in a weakly nonlinear regime driven by the balance of the wave drive and dissipation until reaching saturation. The strongly nonlinear transition is sufficiently fast that it can be treated as occurring instantaneously, and by considering a time-local approximation for the distribution after the flattening has occurred, we find a closed-form analytical solution for the mode amplitude in the weakly nonlinear phase.  A compact piecewise-continuous solution for the entire time evolution of the mode amplitude is therefore constructed.  This result is shown to agree closely with nonlinear kinetic simulations and is derived within a framework common to other physical systems such as galactic discs and viscous fluids.
\end{abstract}

\section{Introduction}

Kinetic instabilities develop as a result of free energy exchange between phase-space gradients in a system's distribution function and a resonant wave field \cite{landau1946,penrose1960}. The strength of these instabilities is characterized by a linear mode growth rate ($\gamma_L$) proportional to the resonant particle distribution gradient in action space \cite{kaufman1972}. Nonlinear saturation occurs via the relaxation of the distribution function to a marginally stable state \cite{oneil1965, mazitov1966} and, in collisional and dissipative systems, via the balance of the energy sources and sinks available to the mode \cite{berk1992,breizman2025}. 

The growth and saturation of kinetic instabilities can have global impacts on the dynamics of systems in which they occur. In fusion plasmas, kinetic excitation of Alfv\'en eigenmodes (AEs, discrete normal modes of the thermal plasma) by fast injected or fusion-born particles can expel energetic particles (EPs) from the device, damaging the walls and degrading confinement \cite{zhang2015,salewski2025}. Excitation of AEs is also observed to suppress turbulence through the generation of zonal modes \cite{du2025}, which may improve confinement. In astrophysical plasmas, kinetic instabilities govern particle acceleration in radiation belts \cite{horne2005} and shocks \cite{axford1969}, and the formation of the magnetopause boundary layer \cite{tsurutani1981}. In self-gravitating systems, kinetic instabilities control the formation of spiral structures in galactic discs \cite{hamilton2024} and the resonant relaxation of dark matter \cite{hamilton2023}. These instabilities share a common mathematical structure across diverse physical systems, motivating the development of theoretical frameworks with broad applicability.  

In general, kinetic instabilities are non-perturbative, meaning that the mode structure and amplitude must be determined self-consistently with the evolution of the distribution function.  However, in cases where the particle drive for the wave comes from a minority population, the problem may be treated perturbatively under the assumption that $\gamma/\omega \ll 1$, where $\gamma$ is the mode net growth rate and $\omega$ is the frequency of the mode. Perturbatively driven kinetic instabilities have been studied extensively near marginal stability, where the linear growth rate of the mode is comparable to its background damping rate. This regime is relevant in systems with only slightly unstable distributions, or in cases where the unstable distribution forms more slowly than the instability grows. Refs. \cite{berk1996, berk1997a, breizman1997} developed an analytic framework to describe the nonlinear evolution of the mode amplitude for a marginally stable instability driven by a minority population for Vlasov-Maxwell systems with an integrable unperturbed Hamiltonian. Their 1D model has been successfully fit to agree with experimental measurements of the evolution of AE amplitudes and frequencies in tokamak plasmas \cite{wong1997,fasoli1998}. Many other studies have also investigated frequency chirping, quasilinear theory, and other phenomena in this regime, e.g., Refs. \cite{berk1997,pinches2004,duarte2017,duarte2019,qu2025}. 

If the linear growth rate of the mode ($\gamma_L$) is much larger than its background damping rate ($\gamma_d$), the wave amplitude and the perturbation to the distribution function are much larger than in the marginally stable regime.  This strongly driven regime is observed in AEs in tokamaks, for example, when sawtooth crashes, involving rapid fluxes of plasma energy out of the core, cause changes to the resonance conditions or relaxation of the EP distribution faster than the mode growth rate, causing very rapid growth of the mode amplitude \cite{fredrickson2000,fredrickson2000a,kramer2001,sharapov2013,ruizruiz2025}. This regime is also relevant outside tokamak plasmas. For example, in galactic discs, interactions between perturbations to the gravitational potential and resonant stars can strongly drive the growth of spiral instabilities \cite{hamilton2024}. Significant progress has been made in theoretically describing the instantaneous rate of power exchange of a mode with a distribution function in the later stages of a strongly driven interaction \cite{zakharov1963} and in finding the amplitude saturation levels of a strongly driven mode \cite{berk1990,berk1990b,berk1990c,petviashvili1999}, but a theoretical description of the time evolution has not yet been developed.  These works provide a foundation on which to develop an analytical model describing the nonlinear time evolution of instabilities driven perturbatively in this regime, which is the goal of this paper.  

We consider the near-resonance interaction of an electrostatic wave with an unstable plasma distribution in one dimension, which is a simplified structure that captures many fundamental properties of these instabilities \cite{berk1997a}. The governing equations are isomorphic to those describing kinetic instabilities in many other systems (described in detail in Section \ref{sec: appl}). In cases that reach steady-state saturation levels, the mode is found to first evolve linearly until it reaches sufficient amplitude, at which point a fast, strongly nonlinear transition occurs in which the slope of the distribution is completely flattened around the resonance.  The system then evolves to saturation in a weakly nonlinear regime.  The strongly nonlinear transition occurs sufficiently fast that it can be considered to be instantaneous, and by considering a time-local approximation for the distribution after the flattening has occurred, we find a closed-form analytical solution for the mode amplitude in the weakly nonlinear phase.  The linear and weakly nonlinear phases are then connected to formulate a piecewise-continuous analytical solution for the time evolution of the mode amplitude, which shows excellent agreement with nonlinear kinetic simulations.  

\section{Reduced kinetic framework} \label{sec: framework}
To model a strongly driven kinetic instability with perturbative drive, we consider the interaction of an externally imposed electrostatic wave with a distribution function with a positive velocity-space gradient in one dimension. Though this may seem like an overly simplistic treatment, the temporal structure of this model is fundamentally the same as that which considers an arbitrarily polarized wave in an axisymmetric tokamak \cite{berk2012}, as well as many systems outside plasma physics (refer to Section \ref{sec: appl} for details), and provides significant insight into the fundamental properties of these resonant interactions. 

The wave electric field is assumed to be of the form $\mathcal{E}(t,x) =  \frac{1}{2}[\tilde{\mathcal{E}}(t)\exp(ikx-i\omega t) + \mathrm{c.c.}]$, where $k$ is the wavenumber, $\omega$ is the wave frequency, and $\tilde{\mathcal{E}}(t) = |\tilde{\mathcal{E}}(t)|e^{i\phi(t)}$ is the slowly varying complex wave amplitude. The exchange of energy between the wave field, the EPs, and the dissipative background is then given by the power balance equation 
\begin{equation}\label{eq: amplitude_x_v}
    \frac{\partial \tilde{\mathcal{E}}}{\partial t} +\gamma_d \tilde{\mathcal{E}} = - 2 q k \int_0^{2\pi/k} \mathrm{d}x \,e^{-i(kx-\omega t)} \int_{-\infty}^{\infty} \mathrm{d}v\, v \mathsf{F}(t,x,v),
\end{equation}
where $q$ is the species charge and $\mathsf{F}(t,x,v)$ is the EP distribution function. The background dissipation rate, $\gamma_d$, is taken to be constant, and models wave damping due to interaction with the continuum, radiative processes, and Landau damping on the thermal bulk plasma. For a derivation of Eq. \ref{eq: amplitude_x_v} in the 1D electrostatic case, see Appendix \ref{appendix: amplitude equation derivation}, and for derivations of this equation for the more general case of arbitrarily polarized waves in 3D, see Refs. \cite{berk1997a,berk2012}. The resonant EP distribution is described by the kinetic equation 
\begin{equation} \label{eq: kinetic vxt}
    \frac{\partial \mathsf{F}}{\partial t}+v\frac{\partial \mathsf{F}}{\partial x}+\frac{q}{m}|\tilde{\mathcal{E}}(t)|\cos[kx-\omega t+\phi(t)]\frac{\partial \mathsf{F}}{\partial v}=C[\mathsf{F}] + S(v),
\end{equation}
where $|\tilde{\mathcal{E}}(t)|$ denotes the real part of the complex field amplitude, $C[\mathsf{F}]$ is the collision operator, and $S(v)$ is a particle source term. The collision operator is taken to describe diffusion in velocity space and is given by $C[\mathsf{F}]=({\nu}_{\text{eff}}^3/k^2)\partial_v^2\mathsf{F}$, where ${\nu}_{\text{eff}}$ is the effective scattering rate \cite{duarte2017b}. By solving Eq. \ref{eq: kinetic vxt} for the equilibrium distribution $\mathsf{F}_0$ with no mode, the source term is found to be given by 
\begin{equation}
     S(v) = -\frac{{\nu}^3_{\text{eff}}}{k^2}\frac{\partial^2\mathsf{F}_0}{\partial v^2}.
\end{equation}
The linear growth rate of a perturbation to this system is given by 
\begin{equation}
    \gamma_L = \frac{2\pi^2 q^2 \omega}{mk^2}\mathsf{F}_0'(v=\omega/k)
\end{equation}
where $\mathsf{F}_0'(v=\omega/k)$ is velocity-space gradient of the unperturbed equilibrium distribution (which does not evolve in time) evaluated at the resonant velocity. It is convenient to move to the reference frame of the wave and write Eqs. \ref{eq: amplitude_x_v} and \ref{eq: kinetic vxt} in dimensionless form. The dimensionless spatial coordinate (angle) and velocity (action) are defined in the reference frame of the wave by $\xi = kx - \omega t$ and $\Omega=k(v-\omega/k)/\gamma_L$, respectively, and the dimensionless mode amplitude is defined by $A(t) = \omega_b^2/\gamma_L^2 = qk|\mathcal{E}|/(m\gamma_L^2)$, where $\omega_b(t)$ is the bounce frequency of the deeply trapped particles (see Appendix \ref{appendix: bounce frequency}).  The dimensionless distribution function is given by $f = 2\pi q^2 \omega/(mk\gamma_L^2) \mathsf{F}$, and time by $t' = \gamma_L {t}$. Under these normalizations and the assumption that the integral $\int \mathrm{d}v\, vf(t,x,v) \simeq (\omega/k)\int \mathrm{d}v\, f(t,x,v)$ near the resonant velocity, the system of equations describing the resonant distribution function and the wave amplitude are written  

\begin{equation}\label{eq: kinetic_Omega_xi}
    \frac{\partial f}{\partial t'}+\Omega\frac{\partial f}{\partial\xi}+|A(t')|\cos(\xi+\phi) \frac{\partial f}{\partial\Omega}=\hat{\nu}_{\text{eff}}^3\frac{\partial^2(f - F_0)}{\partial \Omega^2}
\end{equation} 

\begin{equation}\label{eq: amplitude-Omega_xi}
    \frac{\mathrm{d}A}{\mathrm{d}t'}+\hat{\gamma}_d A=-\frac{1}{\pi}\int_{-\pi/2}^{3\pi/2}\mathrm{d}\xi\, e^{-i\xi}\int_{-\infty}^{\infty}\mathrm{d}\Omega\, f(t',\xi,\Omega), 
\end{equation}
where $\hat{\gamma}_d \equiv \gamma_d/\gamma_L$ and $\hat{\nu}_{\text{eff}} \equiv \nu_{\text{eff}}/\gamma_L$. For simplicity, the calculation that follows is done assuming $q>0$, and the bounds of the integral over $\xi$ coincide with the X-points of the resonant island in that case, but the end result is agnostic to the species charge. The resonance is chosen to be in a region where the slope of the unperturbed equilibrium, $F_0$, is positive, and the width of resonance in velocity space is assumed to be narrow compared to the scale length of the equilibrium distribution function. Therefore, in a narrow range of velocities around the resonance, the equilibrium distribution has an approximately constant slope and can be written in the dimensionless coordinates as $F_0 = (\omega + \Omega)/\pi$. 

It should be noted that one can obtain an identical amplitude equation from the Vlasov-Possion system under the assumption of weak wave drive, as is shown in Ref. \cite{del-castillo-negrete1998}. Eq. \ref{eq: amplitude-Omega_xi}, however, is derived using only the WKB approximation ($\gamma/\omega \ll 1$, see Appendix \ref{appendix: amplitude equation derivation}), and is valid for both weakly and strongly driven instabilities provided that this ordering is not violated.  

\subsection{Equations for the mode amplitude and phase}
It will be convenient to work in the coordinates $\tau = t'$, $z = \xi+\phi(t')$, and $y = \Omega^2/(2|A(t')|)-\sin[\xi+\phi(t')]$, where $y$ is the particle energy normalized by the wave amplitude. In these coordinates, Eq. \ref{eq: amplitude-Omega_xi} is written
\begin{equation}
\frac{\mathrm{d}A}{\mathrm{d}\tau}+\hat{\gamma}_{d}A=-\frac{\sqrt{|A|}}{\pi\sqrt{2}}\int_{-\pi/2+\phi}^{3\pi/2+\phi}\mathrm{d}z\, e^{-i(z-\phi)}\int_{-\sin z}^{\infty}\frac{\mathrm{d}y}{\sqrt{y+\sin z}}(f^+ +f^-).
\end{equation}
Writing the complex amplitude as $A(\tau) = |A(\tau)|e^{i\phi(t)}$, this becomes 

\begin{equation}\label{eq: amplitude_comp}
\frac{\mathrm{d}|A|}{\mathrm{d}\tau}+i|A|\frac{\mathrm{d}\phi}{\mathrm{d}\tau}+\hat{\gamma}_{d}|A|=-\frac{\sqrt{|A|}}{\pi\sqrt{2}}\int_{-\pi/2+\phi}^{3\pi/2+\phi}\mathrm{d}z\int_{-\sin z}^{\infty}\frac{\mathrm{d}y\, e^{-iz}}{\sqrt{y+\sin z}}(f^+ +f^-).
\end{equation}
The real and imaginary parts of Eq. \ref{eq: amplitude_comp} give equations for the magnitude and phase of the complex amplitude:

\begin{equation}\label{eq: amplitude_z_E_amp}
\frac{\mathrm{d}|A|}{\mathrm{d}\tau}+\hat{\gamma}_{d}|A|=-\frac{\sqrt{|A|}}{\pi\sqrt{2}}\int_{-\pi/2+\phi}^{3\pi/2+\phi}\mathrm{d}z\int_{-\sin z}^{\infty}\frac{\mathrm{d} y\, \cos z}{\sqrt{y+\sin z}}(f^+ +f^-) 
\end{equation}

\begin{equation}\label{eq: amplitude_z_E_phase}
\frac{\mathrm{d}\phi}{\mathrm{d}\tau}=\frac{1}{\pi\sqrt{2|A|}}\int_{-\pi/2+\phi}^{3\pi/2+\phi}\mathrm{d}z\int_{-\sin z}^{\infty}\frac{\mathrm{d} y\, \sin z}{\sqrt{y+\sin z}}(f^+ +f^-).
\end{equation}
The phase $\phi$ describes any change to the wave frequency occurring during the evolution. 
\section{Distribution function}\label{sec: distribution function}
To calculate the mode evolution using Eqs. \ref{eq: amplitude_z_E_amp} and \ref{eq: amplitude_z_E_phase}, the EP distribution function, $f$, must be determined locally around the resonance. In the strongly driven regime, the gradient in the distribution is sufficiently positive that the linear growth rate of the mode far exceeds the damping rate, i.e., $\gamma_L \gg \gamma_d$. The mode evolution is determined by the balance of three processes: (i) phase mixing (occurring at a rate roughly proportional to $\omega_b$), which controls the amount of energy the wave gains from the distribution per unit time (ii) scattering (occurring at a rate $\nu_{\text{eff}}$), which controls how much energy enters the resonance from the upstream distribution per unit time, and (iii) damping, in which the mode loses energy a rate $\gamma_d$.

The evolution of the mode in this regime can be schematically understood to occur in three parts.  From a noise amplitude level, the mode first grows linearly at a rate $\gamma_L$.  Linear growth continues as long as ${\nu}_{\text{eff}}^3\gg \omega_b^3$, i.e., the rate at which scattering reinforces the distribution gradient exceeds the rate at which phase mixing can flatten the gradient.  Although the mode amplitude is growing, the distribution in this phase will be unchanged from the unperturbed equilibrium, and the net growth rate is therefore given by $\gamma_L-\gamma_d$.  When the mode amplitude (which is $\propto \omega_b^2$) grows to a sufficient level that $\omega_b^3 \sim {\nu}_{\text{eff}}^3$, phase mixing occurs fast enough that the distribution function flattens around the resonance.  This is a strongly nonlinear process, and cannot be described by the model presented here, but it occurs on a timescale ($\sim \omega_b^{-1}$) much faster than the timescale on which the mode reaches its nonlinear saturation, i.e., $\omega_b \gg \gamma$, where $\gamma = \gamma_{NL}(\tau)-\gamma_d$ is the net growth rate and $\gamma_{NL}(\tau)$ is the nonlinear growth rate due to the EP drive. After this brief strongly nonlinear transition, the mode amplitude is large enough that $\omega_b^3 \gg {\nu}_{\text{eff}}^3$, and the system evolves in a weakly nonlinear regime until it reaches its saturation level, which is determined by the balance of dissipation ($\gamma_{d}$) and energy from new particles entering the resonance via scattering ($\nu_{\text{eff}}$).  This saturation level has been predicted for several collision operators by Refs. \cite{berk1990,berk1990b,berk1990c,petviashvili1999}.  

\begin{figure}[h]    
    \begin{center}  
    \includegraphics[width=0.45\textwidth]{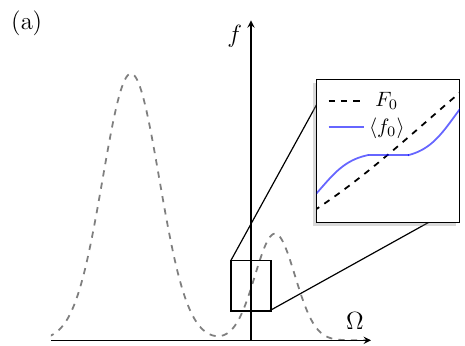} 
    \hspace{1cm}
    \includegraphics[width=0.4\textwidth]{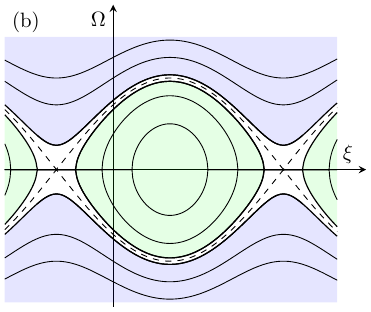} 
    \end{center} 
    \caption{(a) A schematic bump-on-tail distribution (dashed black), where the resonance is located at $\Omega=0$. The inset shows the unperturbed equilibrium, $F_0$ in dashed black and the $\xi$-averaged 0$^{\text{th}}$ order distribution during the weakly nonlinear phase, $\langle f_0 \rangle$, in blue. (b) Contours of constant energy plotted in $(\Omega,\xi)$ phase space.  The passing particle regions are shaded in blue, and the trapped particle regions in green. The separatrix is shown by the black dashed line, and the boundary layer is shown by the white region around it.} \label{fig: phase space}
\end{figure} 

Since the strongly nonlinear phase is very short compared to the timescale on which the mode reaches saturation, we hypothesize that the time evolution of the mode amplitude can be modeled by considering only the linear phase and the weakly nonlinear phase.  In the linear phase, which will be referred to as Phase I, the evolution is described by the exponential growth at the net linear growth rate which depends on the unperturbed equilibrium $F_0(\Omega)$.  In the weakly nonlinear phase, which will be referred to as Phase II, the evolution is described by perturbative dynamics occurring on top of the distribution after the strongly nonlinear phase, $f_0$.  $f_0$ is the equilibrium distribution in the presence of the mode and the absence of collisions and dissipation, where the velocity-space gradient has been completely flattened around the resonance, and it is allowed to evolve in time through the wave amplitude.  A schematic plot of these distributions in velocity space is shown in the inset of Fig. \ref{fig: phase space}(a).  In the following sections, we will develop a theoretical framework with which to describe the dynamics in the weakly nonlinear Phase II of the evolution.

In the absence of collisions, there are two classes of particles in the distribution function:  ``trapped'' particles, which bounce back and forth in the potential well of the wave, and ``passing'' particles whose trajectories are deflected by the wave but which do not bounce. These two populations are separated by the separatrix. In the regime where $\nu_{\text{eff}}^3/\omega_b^3 \ll 1$, collisions have a weak influence on the distribution function away from the separatrix. These particles will still be referred to as belonging to ``trapped'' and ``passing'' parts of the distribution function. Near the separatrix, a thin boundary layer exists where collisions have a more dominant effect on the dynamics. This boundary layer will be discussed in more detail in a later section. The trapped, passing, and boundary layer phase-space regions are shown  Fig. \ref{fig: phase space}(b).

\subsection{Local transport structure at zeroth order}
Again, it is convenient to introduce the coordinates $\tau = t'$, $z = \xi+\phi(t')$, and $y = \Omega^2/(2|A(t')|)-\sin[\xi+\phi(t')]$, in which the kinetic equation (Eq. \ref{eq: kinetic_Omega_xi}) is given by
\begin{multline}\label{eq: kinetic_y_z}
    \frac{\partial f^{\pm}}{\partial \tau} + \left[\phi'(\tau) \pm  \sqrt{2|A(\tau)|}\sqrt{y+\sin z}\right]  \frac{\partial f^{\pm}}{\partial z}\\-\left[\frac{|A'(\tau)|}{|A(\tau)|}(y+\sin z)+\phi'(\tau)\cos z \right]\frac{\partial f^{\pm}}{\partial y}  = \frac{2\hat{\nu}_{\text{eff}}^3}{|A(\tau)|} \sqrt{y+\sin z}\frac{\partial}{\partial y}\sqrt{y+\sin z} \frac{\partial f^{\pm}}{\partial y},
\end{multline}

where the upper and lower signs refer to particles with $\Omega>0$ and $\Omega<0$, respectively. Refer to Appendix \ref{appendix: coordinate transform} for the detailed transformation between the two coordinate systems. Away from the separatrix, the distribution function can be expanded as $f \simeq f_0 + f_1$, where $f_1 \sim \epsilon (f_0-F_0)$ and $\epsilon \sim \nu_{\text{eff}}^3/\omega_b^3  \sim \gamma/\omega_b \ll 1$. Note that for now, no subsidiary ordering of these two small parameters is considered and they will both appear to first order, although they are not necessarily of comparable magnitude at all times. To $\mathcal{O}(1)$, Eq. \ref{eq: kinetic_y_z} is simply   
\begin{equation} \label{eq: f0_zeorth_order}
    \pm \frac{\partial f_0^{\pm}}{\partial z}  = 0, 
\end{equation}
which indicates that the zeroth order distribution is $f^\pm_0 = f^\pm_0(\tau,y)$, a function only of time and the particle energy.  To $\mathcal{O}(\epsilon)$, Eq. \ref{eq: kinetic_y_z} is given by
\begin{multline}\label{eq: kinetic_first_order}
    \frac{\partial f_0^{\pm}}{\partial \tau} +\phi'(\tau) \frac{\partial f_0^{\pm}}{\partial z} \pm  \sqrt{2|A(\tau)|}\sqrt{y+\sin z} \frac{\partial f_1^{\pm}}{\partial z}\\-\left[\frac{|A'(\tau)|}{|A(\tau)|}(y+\sin z)+\phi'(\tau)\cos z \right]\frac{\partial f_0^{\pm}}{\partial y} = \frac{2\hat{\nu}_{\text{eff}}^3}{|A(\tau)|} \sqrt{y+\sin z}\frac{\partial}{\partial y}\sqrt{y+\sin z} \frac{\partial f_0^{\pm}}{\partial y},
\end{multline}
To gain some physical intuition for the transport characteristics at zeroth order, we consider now only the passing particles.  Enforcing that $f_1$ be $2\pi$-periodic in the angle $z$ and taking the angle average, defined by $\langle f(\tau, z, y)\rangle = \int_{-\pi/2}^{3\pi/2}f(\tau, z, y)/2\pi\, \mathrm{d}z$, we find that $f_0$ satisfies the equation
\begin{eqnarray} \label{eq: transport_advdiff}
     \left \langle \frac{1}{\sqrt{y+\sin z}}\right \rangle \frac{\partial f_0^{\pm(\mathrm{p})}}{\partial \tau} - \left[\frac{|A'(\tau)|}{|A(\tau)|}\left \langle \sqrt{y+\sin z} \right \rangle  \right] \frac{\partial f_0^{\pm(\mathrm{p})}}{\partial y}  =\frac{2\hat{\nu}_{\text{eff}}^3}{|A(\tau)|} \frac{\partial}{\partial y}\left \langle\sqrt{y+\sin z}\right \rangle \frac{\partial f_0^{\pm(\mathrm{p})}}{\partial y},
\end{eqnarray}
where the superscript $(\mathrm{p})$ denotes the distribution for passing particles. The term involving $\phi'(\tau)$ vanishes because its integral coefficient $\left \langle \cos z/\sqrt{y+\sin z} \right \rangle =0$. Eq. \ref{eq: transport_advdiff} is an advection-diffusion equation in one dimensional phase space (as opposed to the two dimensional phase space of Eq. \ref{eq: kinetic_y_z}). This reduction of dimension is indicative of the fact that particles move along constant energy surfaces in phase space very fast compared to collisional and mode growth timescales.  Transport occurs in energy space by both advection by KAM energy surfaces as they expand and contract due to the changing wave amplitude and by scattering of particles from one energy surface to another.

In order to make further progress, some subsidiary ordering of the two small parameters $\nu_{\text{eff}}^3/\omega_b^3$ and $\gamma/\omega_b$ must be assumed. The relation between these two parameters describes the relative importance of collisions and dissipation to the dynamics. For sufficiently slow collisions ($\nu_{\text{eff}}^3/\omega_b^3 \ll \gamma/\omega_b$), the right-hand side of Eq. \ref{eq: transport_advdiff} can be neglected, and the transport is solely advective. In the case of steady-state saturation levels, this is unlikely to be a dominant regime unless $\nu_{\text{eff}}\ll\gamma$ even within the narrow resonance region, as the net growth rate $\gamma$ becomes vanishingly small as the mode reaches saturation. In the limit where the effective collisional timescale is sufficiently fast compared to the net growth rate of the mode ($\nu_{\text{eff}}^3/\omega_b^3 \gg \gamma/\omega_b$), the system is collisional enough to erase particle phase-space correlations within the amplitude evolution timescale.  This is automatically satisfied as the mode approaches saturation in the case of a steady state solution, since $\gamma \rightarrow 0$, and allows the time derivative terms of Eq. \ref{eq: kinetic_y_z} to be neglected. 

\subsection{Asymptotic solutions to the distribution function} \label{sec: asymptotic solutions}
We hypothesize that the dynamics in the weakly nonlinear phase (Phase II) for steady saturation levels can be approximated by taking $\epsilon \equiv \nu_{\text{eff}}^3/\omega_b^3 \gg \gamma/\omega_b$.  This constitutes an assumption of time-locality or Markovian evolution; the distribution function will evolve in time only through the evolution of the mode amplitude. Therefore terms of $\mathcal{O}(\gamma/\omega_b)$ are neglected to all orders and the kinetic equation becomes
\begin{equation}\label{eq: kinetic_y_z_nodt}
    \pm  \frac{\partial f^{\pm}}{\partial z}  = \sqrt{2}\frac{\hat{\nu}_{\text{eff}}^3}{|A(\tau)|^{3/2}} \frac{\partial}{\partial y}\sqrt{y+\sin z} \frac{\partial f^{\pm}}{\partial y}.
\end{equation}

This assumption will restrict the regimes of validity of this model to those with sufficiently large $\nu_{\text{eff}}$ and sufficiently small $\gamma = \gamma_{NL}-\gamma_d$ for the relation $\nu_{\text{eff}}^3/\omega_b^3 \gg \gamma/\omega_b$ to hold. There therefore are two categories of solutions which will not be predicted by this model, both involving non-steady state solutions. The first is those cases with saturated states with amplitudes which oscillate around some constant value, meaning that $\gamma$ never vanishes.  This behavior is associated with self-sustaining temporal feedback between the distribution function and the wave, and this model will predict only the average amplitude level in those cases.  The second is frequency chirping, which is associated with the formation of phase-space holes and clumps at the edges of the resonant island where phase mixing leaves density perturbations that can become ``untrapped'' by the wave in the case of sufficient wave dissipation \cite{bierwage2021,lilley2014}. These sideband perturbations eventually evolve into chirping modes, which are BGK waves \cite{bernstein1957} with frequencies slightly upshifted and downshifted from that of the original wave.  Chirping has been found to occur in the strongly driven regime \cite{lilley2014}, but this requires sufficiently weak collisionality such that dissipation still plays a dominant role, which would imply $\nu_{\text{eff}}^3/\omega_b^3 \sim \gamma/\omega_b$. 

\begin{figure}[h]    
    \begin{center}  
    \includegraphics[width=0.45\textwidth]{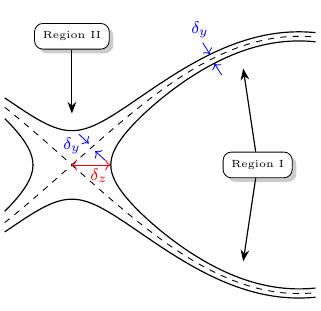}  
    \end{center}
    \caption{Boundary layer regions.  Region I is the part of the boundary layer away from the X-points and Region II is the part of the boundary layer immediately surrounding the X-points.} \label{fig: BL regions}
\end{figure}

The structure of Eq. \ref{eq: kinetic_y_z_nodt} is common to other problems involving particle trapping and a detailed treatment of a very similar boundary layer problem is presented in Ref. \cite{chandran1999}. Away from the separatrix, particles are classified as either trapped or passing. Near the separatrix ($y=1$), the derivatives on the right hand side become large, and there is a boundary layer of width $\delta_y \sim \epsilon^{1/2}$. Near the X-points, which are located at $z=-\pi/2$ and $z=3\pi/2$, the form of $y = \Omega^2/(2|A(t')|)-\sin [\xi+\pi(t')]$ requires another layer of width $\delta_z\sim \epsilon^{1/4}$ in $z$. These constitute two regions within the boundary layer which must be considered: Region I, away from the X-points (where $y+ \sin z \sim 1$), and Region II around the X-Points (where $y+ \sin z \ll 1$). These regions are shown in Fig. \ref{fig: BL regions}. 

Up to $\mathcal{O}(\epsilon)$ contributions to the amplitude and phase equations will be considered. This requires calculating up through $\mathcal{O}(\epsilon)$ in the outer solution.  The contribution from Region I of the boundary layer is at most $\mathcal{O}(\epsilon)$, so only zeroth order is needed.  The contribution from Region II of the boundary layer is at most $\mathcal{O}(\epsilon^{5/4})$, which again only requires the zeroth order solution. $\epsilon$ shall be considered to be sufficiently small that any term of higher order than $\mathcal{O}(\epsilon)$ can be neglected, and so Region II will not contribute to the amplitude and phase equations. 

\subsubsection{Outer solution}
Away from the separatrix, Eq. \ref{eq: kinetic_y_z_nodt} is expanded in powers of $\epsilon = \nu_{\text{eff}}^3/\omega_b^3$. To $\mathcal{O}(1)$ the solution to Eq. \ref{eq: kinetic_y_z} is again simply $f^\pm_0 = f^\pm_0(\tau,y)$.  To $\mathcal{O}(\epsilon)$, Eq. \ref{eq: kinetic_y_z} becomes

\begin{equation} \label{eq: kinetic_f1}
     \frac{\partial f_1^{\pm(\mathrm{o})}}{\partial z}  = \pm \sqrt{2}\frac{\hat{\nu}_{\text{eff}}^3}{|A(\tau)|^{3/2}}\frac{\partial}{\partial y}\sqrt{y+\sin z} \frac{\partial f_0^{\pm(\mathrm{o})}}{\partial y},
\end{equation}
where the superscript $(\mathrm{o})$ denotes the ``outer" solution. The form of $f_0^{(\mathrm{o})}$ can be found using solvability conditions on Eq. \ref{eq: kinetic_f1}.  For passing particles, which have $y>1+\epsilon^{1/2}$, we enforce that $f_1$ be $2\pi$-periodic in the angle $z$ (so that it cannot grow without bound) and take the angle average, defined by $\langle f(\tau, z, y)\rangle \equiv \int_{-\pi/2}^{3\pi/2}f(\tau, z, y) \, /2\pi \mathrm{d}z$. This results in 
\begin{equation} \label{eq: kinetic_f0}
    0  = \pm \frac{\partial}{\partial y}\left \langle \sqrt{y+\sin z} \right \rangle \frac{\partial f_0^{\pm(\mathrm{p})}}{\partial y},
\end{equation}
where the superscript $(\mathrm{p})$ denotes passing particles. Integrating once over $y$, and enforcing that $\lim_{y\rightarrow \infty} \partial_y f_0^{\pm(\mathrm{p})} = \lim_{y\rightarrow \infty} \partial_y F_0(y)$ one finds
\begin{equation} \label{eq:f_0 passing derivative}
    \frac{\partial f_0^{\pm(\mathrm{p})}}{\partial y} = \frac{\pm \sqrt{2|A(\tau)|}}{\int_{-\pi/2}^{3\pi/2} \mathrm{d}z\, \sqrt{y+\sin z}}.
\end{equation}

For trapped particles, which have $-1<y<1-\epsilon^{1/2}$, Eq. \ref{eq: kinetic_f1} can be integrated between the bounce points, $z_1$ and $z_2$ to find 
\begin{equation} 
    f^{\pm (\mathrm{t})}_1(\tau,z_1,y) - f^{\pm (\mathrm{t})}_1(\tau,z_2,y) = \pm \sqrt{2}\frac{\hat{\nu}_{\text{eff}}^3}{|A(\tau)|^{3/2}}\frac{\partial}{\partial y}\int_{z_1}^{z_2} \mathrm{d}z\, \sqrt{y+\sin z} \frac{\partial f_0^{\pm(\mathrm{t})}}{\partial y}. 
\end{equation} 
Subtracting the plus and minus equations from one another and enforcing continuity at the bounce points such that $f^+_1(\tau,z_{1,2},y) =f^-_1(\tau, z_{1,2},y) $ yields 
\begin{equation} 
    0 = \sqrt{2}\frac{\hat{\nu}_{\text{eff}}^3}{|A(\tau)|^{3/2}}\frac{\partial}{\partial y}\int_{z_1}^{z_2} \mathrm{d}z\,\sqrt{y+\sin z} \left( \frac{\partial f_0^{+(\mathrm{t})}}{\partial y} +\frac{\partial f_0^{-(\mathrm{t})}}{\partial y}\right). 
\end{equation} 
Integrating once over $y$, enforcing that $\partial_y f_0^{\pm(\mathrm{t})}$ be finite at the O-point, where $\sqrt{y+\sin z} = 0$ and assuming no particle source at the O-point, gives
\begin{equation} \label{eq:f_0 trapped derivative}
    \frac{\partial f_0^{\pm(\mathrm{t})}}{\partial y} = 0, 
\end{equation}
implying that to zeroth order, the trapped particle distribution is completely flattened.  These zeroth order distributions can be written in terms of the complete elliptic integral of the second kind, $E(k)$, 
\begin{equation}\label{eq:f_0 elliptic derivative}
    \frac{\partial f_0^{\pm(\mathrm{o})}}{\partial y} =
    \begin{cases}
        \displaystyle 0, & -1 \leq y \leq 1-\epsilon^{1/2}\\
        \displaystyle \pm \frac{k\sqrt{|A(\tau)|}}{4E(k)}, & y\geq 1+\epsilon^{1/2}
    \end{cases}
\end{equation}
where $k^2(y) = 2/(1+y)$. Note that Eqs. \ref{eq: kinetic_f1} and \ref{eq:f_0 elliptic derivative} will be sufficient for the calculation of the real amplitude in Eq. \ref{eq: amplitude_z_E_amp}.  The forms of $f_0^{\pm(\mathrm{o})}$ and $f_1^{\pm(\mathrm{o})}$ are needed for the calculation of the phase in \ref{eq: amplitude_z_E_phase}. 

To find the form of $f_0^{\pm(\mathrm{o})}$, Eq. \ref{eq:f_0 elliptic derivative} is integrated over $y$.  The trapped particle distribution, $f_0^{(\mathrm{t})}$, must be unchanged from $F_0$ at the O-point ($y = -1$), and the discontinuity across the boundary layer is assumed to be of $\mathcal{O}(\epsilon^{1/2})$. The zeroth order distribution function outside of the separatrix boundary layer is then given by 
\begin{equation}\label{eq:f_0}
    f_0^{\pm(\mathrm{o})}(\tau,y) =
    \begin{cases}
        \displaystyle \frac{\omega}{\pi}, & -1 \leq y \leq 1-\epsilon^{1/2}\\
        \displaystyle \frac{\omega}{\pi} \pm \int_{1}^y \mathrm{d}y' \frac{k(y')\sqrt{|A(\tau)|}}{4E(k(y'))}, & y\geq 1+\epsilon^{1/2}.
    \end{cases}
\end{equation} 
To next order, Eqs. \ref{eq:f_0 passing derivative} and \ref{eq:f_0 trapped derivative} are substituted into Eq. \ref{eq: kinetic_f1}, which can be integrated over $z$. For the trapped particles, this yields
\begin{equation} \label{eq: f1 trapped}
    f_1^{\pm(\mathrm{t})}(\tau,z,y)  = c_1^{\pm(\mathrm{t})}(\tau,y),
\end{equation} 
and for the passing particles this yields 
\begin{equation} \label{eq: f1 passing}
    f_1^{\pm(\mathrm{p})}(\tau,z,y)  = c_1^{\pm(\mathrm{p})}(\tau,y) + \frac{2\hat{\nu}_{\text{eff}}^3}{|A(\tau)|}\frac{\partial}{\partial y} \left(\frac{\int_{z_{1,2}}^{z} \mathrm{d}z'\sqrt{y+\sin z'}}{\int_{-\pi/2}^{3\pi/2} \mathrm{d}z \sqrt{y+\sin z}} \right).
\end{equation}
Note that the lower bound in the integral in the numerator must be taken to be the physical lower bound of the system, which for $q>0$ is $z_1 = -\pi/2$ for particles with $\Omega>0$ and  $z_2 = 3 \pi/2$ for particles with $\Omega<0$. The constants of integration are found to be $c_1^{\pm(\mathrm{t})}(\tau,y)=0$ and $c_1^{\pm(\mathrm{p})}(\tau,y)=0$ using additional solvability conditions on the kinetic equation (refer to Appendix \ref{appendix: f1 integration constants} for the detailed calculations). Eq. \ref{eq: f1 passing} can be written in a closed form in terms of elliptic integrals, which yields for the outer solution
\begin{equation}\label{eq: f1}
    f_1^{\pm(\mathrm{o})}(\tau,z, y) =
    \begin{cases}
        0 & -1 \leq y \leq 1-\epsilon^{1/2}\\
        \begin{aligned}
          &\frac{\hat{\nu}_{\text{eff}}^3k^2}{4|A(\tau)|E(k)^2} \bigg(K(k)[E(\varphi,k)-E(\varphi_{1,2},k)]\\
          &\quad ~~~~~~~~~~- E(k)[F(\varphi,k)-F(\varphi_{1,2},k)]\bigg)
        \end{aligned} & y\geq 1+\epsilon^{1/2}
      \end{cases}
\end{equation}
where $\varphi = \pi/4 - z/2$, $\varphi_1 = 0$, and $\varphi_2 = \pi$, $K(k)$ and $F(\varphi,k)$ are the complete and incomplete elliptic integrals of the first kind, respectively, and $E(\varphi,k)$ is the incomplete elliptic integral of the second kind.  The distribution function is then described outside of the boundary layer by Eqs. \ref{eq:f_0} and \ref{eq: f1}, which depend on the amplitude $|A(\tau)|$, which will be calculated in Section \ref{sec: real amp}.  These results for the distribution function are then compared with simulations in Section \ref{sec: dist evolution}.

\subsubsection{Boundary layer}
In Region I of the boundary layer, we define an ``inner'' layer coordinate by $y = 1+\epsilon^{1/2}\eta$.  To $\mathcal{O}(1)$, the  Region I distribution function, $f^{\pm(\mathrm{iI})}$, satisfies
\begin{equation}\label{eq: reg1_f0}
    \pm  \frac{\partial f_0^{\pm(\mathrm{iI})}}{\partial z}  = \sqrt{2}\sqrt{1+\sin z} \frac{\partial^2 f_0^{\pm(\mathrm{iI})}}{\partial \eta^2}, 
\end{equation} 
where the superscript $(\mathrm{iI})$ denotes the inner solution in Region I of the boundary layer. The contribution to the amplitude integrals from $f_0^{\pm(\mathrm{iI})}$ is at most $\mathcal{O}(\epsilon)$, so no higher order corrections are required. Eq. \ref{eq: reg1_f0} is sufficient to calculate the contribution of Region I to the equations for the amplitude and phase of the wave. 

In Region II, the contribution to the amplitude integrals from the zeroth order distribution is at most $\mathcal{O}(\epsilon^{5/4})$, and therefore it is not necessary to calculate the distribution function in this region (though it can be determined, see Ref. \cite{chandran1999} for details).

\section{Evolution of the real amplitude} \label{sec: real amp}
\subsection{Solution for the real amplitude evolution}
The real part of the amplitude satisfies Eq. \ref{eq: amplitude_z_E_amp}. It is convenient to integrate by parts over $z$ to find  
\begin{equation}\label{eq: amplitude_z_E_amp_int}
    \frac{\mathrm{d}|A|}{\mathrm{d}\tau}+\hat{\gamma}_{d}|A|=\frac{\sqrt{2|A|}}{\pi}\underbrace{\int_{-\pi/2+\phi}^{3\pi/2+\phi} \mathrm{d}z\int_{-\sin z}^{\infty}\mathrm{d}y\, \sqrt{y+\sin z} \frac{\partial}{\partial z}(f^+ +f^-),}_{\mathcal{I}}
\end{equation}
Substituting the forms of the distribution function found in the previous sections and evaluating numerically up to $\mathcal{O}(\epsilon)$, the integral is 
\begin{equation}
    \mathcal{I} \approx 3.90 |A|^{1/2} \epsilon.
\end{equation} 
See Appendix \ref{appendix: real amplitude} for the detailed calculations. Using $\epsilon \equiv \hat{\nu}_{\text{eff}}^3/|A|^{3/2}$, the real amplitude therefore satisfies 
\begin{equation}\label{eq: amplitude_final}
    \frac{\mathrm{d}|A|}{\mathrm{d}\tau}+\hat{\gamma}_{d}|A|= 1.756 \frac{\hat{\nu}_{\text{eff}}^3}{|A|^{1/2}}.
\end{equation}
Eq. \ref{eq: amplitude_final} is a nonlinear Bernoulli ODE, which has the exact analytical solution
\begin{equation}\label{eq: amplitude_solution}
    |A(\tau)| = \left[e^{-3\hat{\gamma}_{d}(\tau-\tau_0)/2}\left(|A_0|^{3/2}-|A_{\text{s}}|^{3/2}\right)+|A_{\text{s}}|^{3/2}\right]^{2/3}
\end{equation} where $|A_{\text{s}}|^{3/2} = 1.756\hat{\nu}_{\text{eff}}^3/\hat{\gamma}_{d} $  (the same final saturation level as that found in Ref. \cite{petviashvili1999}). The amplitude $|A(\tau = \tau_0)|=|A_0|$ is the amplitude at which the strongly nonlinear transition occurs, where $\tau_0$ is the first time at which the fully flattened distribution $f_0$ is established. Empirically, it is found that in cases with weak collisions ($\hat{\nu}_{\text{eff}} < 1 $), $|A_0|$ is very close to the collisionless saturation level, $|A_c| = (3.2)^2 $ \cite{fried1971}. In cases with stronger collisions ($\hat{\nu}_{\text{eff}} \gg 1 $), the linear phase is extended far beyond the collisionless level and $|A_0|$ can be significantly larger then $|A_c|$.  The nonlinear growth rate, $\gamma_{NL}(\tau)$, of the mode during Phase II can be defined via 
\begin{equation}\label{eq: NL_gr_def}
    |A(\tau)| = |A_{0}|\exp \left[\int_{\tau_0}^\tau \mathrm{d}\tau'(\hat{\gamma}_{NL}(\tau')-\hat{\gamma}_d)\right],
\end{equation} resulting in
\begin{equation} \label{eq: gammaNL}
    \hat{\gamma}_{NL}(\tau) =  \hat{\gamma}_{d}\left( \frac{|A_{\text{sat}}|}{|A(\tau)|}\right)^{3/2}, 
\end{equation} where $A(\tau)$ is given by Eq. \ref{eq: amplitude_solution}.

\subsection{Comparison of the predicted amplitude with simulations}
The predictions of Eqs. \ref{eq: amplitude_solution} and \ref{eq: gammaNL} can be directly compared with the results of nonlinear kinetic simulations performed using the Bump On Tail (BOT) code, which numerically solves Eqs. \ref{eq: kinetic_Omega_xi} and \ref{eq: amplitude-Omega_xi} using a Fourier series representation of the distribution function \cite{lilley2010}.
   
\begin{figure}  
\begin{center}
\includegraphics[width=0.9\textwidth]{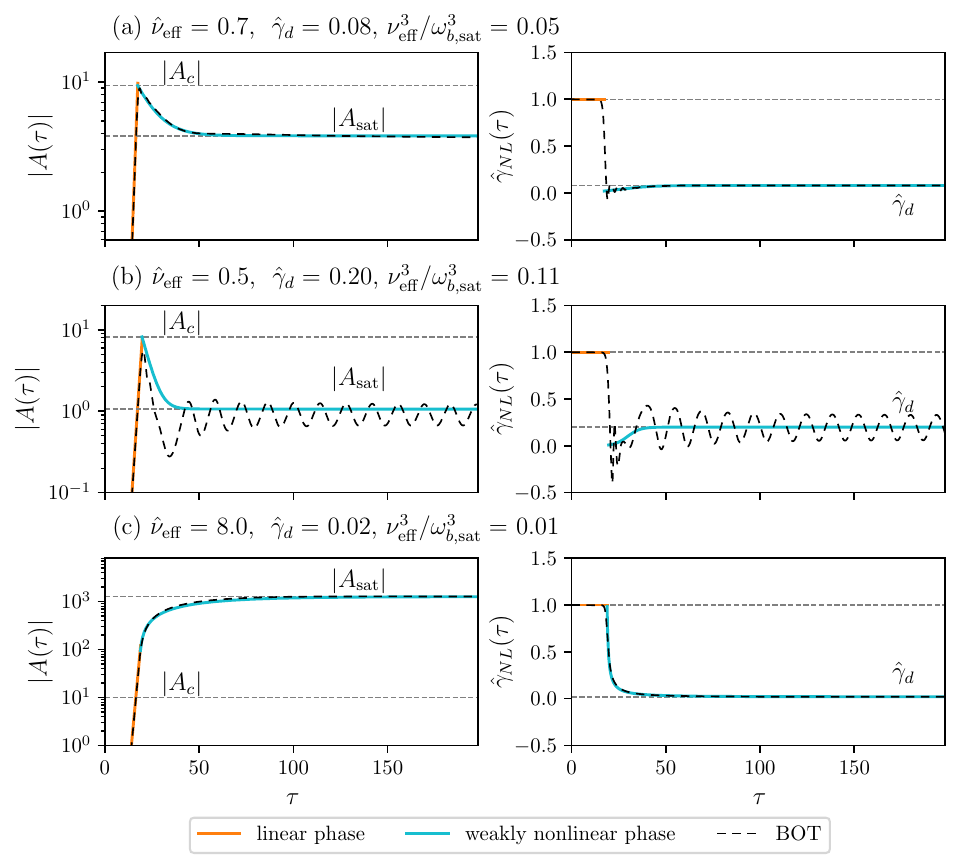}
\end{center}
\caption{Comparison of theoretical predictions of amplitude (left) and nonlinear growth rate (right), Eqs. \ref{eq: amplitude_solution} and \ref{eq: gammaNL}, with nonlinear kinetic simulations (using the BOT code) for three example cases: (a) $\hat{\nu}_{\text{eff}} = 0.7$ and $\hat{\gamma}_d = 0.08$, (b)  $\hat{\nu}_{\text{eff}} = 0.5$ and $\hat{\gamma}_d = 0.2$, and (c) $\hat{\nu}_{\text{eff}} = 8.0$ and $\hat{\gamma}_d = 0.02$.  The amplitude saturation level $|A_{\text{sat}}| = (1.756\hat{\nu}_{\text{eff}}^3/\hat{\gamma}_{d})^{2/3}$ \cite{petviashvili1999}, the collisionless saturation level $|A_c|= (3.2)^2$  \cite{fried1971}, and the final growth rate of $\hat{\gamma}_{NL}=\hat{\gamma}_d$ are shown in dotted black. Phase I (linear phase) is shown in orange and Phase II (weakly nonlinear phase) is shown in blue.}\label{fig: compare_BOT}
\end{figure} 

The left column of Fig. \ref{fig: compare_BOT} shows the predictions of Eq. \ref{eq: amplitude_solution} compared with nonlinear kinetic simulations performed using the BOT code \cite{lilley2010} for several example cases. We construct a piecewise-smooth theoretical prediction for the entire amplitude evolution by stitching together the exponential growth phase, $|A(\tau)| = |A_0| \exp[(1-\hat{\gamma}_d)\tau]$ (in orange), with  Eq. \ref{eq: amplitude_solution} (in blue) at $\tau = \tau_0$. This prediction shows very close agreement with the BOT simulations. In panels (a) and (b), $\hat{\nu}_{\text{eff}} < 1$, and $|A_0| \approx |A_c|$, as shown by the dotted line.  Panel (b) shows a case  with an oscillating saturation, where the theoretical prediction follows approximately the average value of the saturation level; this is a similar result to that found in the marginally stable regime using the time-local approximation \cite{duarte2019a}. Panel (c) shows a case where $\hat{\nu}_{\text{eff}}^3 \gg 1$ and the linear phase is extended far beyond the predicted collisionless level.  The right column of Fig. \ref{fig: compare_BOT} shows a comparison of the BOT growth rate with the linear growth rate (in orange) and Eq. \ref{eq: gammaNL} for the nonlinear growth rate (in blue).

\section{Evolution of the phase} \label{sec: phase}
\subsection{Integration of the phase equation}
The phase satisfies Eq. \ref{eq: amplitude_z_E_phase}. The right hand side can be integrated by parts over $y$ to find 
\begin{equation}
    \frac{\mathrm{d}\phi}{\mathrm{d}\tau}= \frac{\sqrt{2}}{\pi\sqrt{|A|}} \mathcal{J}, 
\end{equation}
where the integral $\mathcal{J}$ is given by 
\begin{equation}
    \mathcal{J} = \int_{-\pi/2+\phi}^{3\pi/2+\phi}\mathrm{d}z\bigg\{\lim_{y\rightarrow \infty}\left[\sin z\sqrt{y+\sin z}(f^+ +f^-)\right] - \int_{-\sin z}^{\infty} \mathrm{d} y\, \sin z\sqrt{y+\sin z} \frac{\partial }{\partial y}(f^+ +f^-)\bigg\}
\end{equation}
Up to $\mathcal{O}(\epsilon)$, $\mathcal{J} \simeq 0$, implying that 
\begin{equation}\label{eq: phase final}
    \frac{\mathrm{d}\phi}{\mathrm{d}\tau}= 0.
\end{equation}
See Appendix \ref{appendix: phase} for the detailed calculations. There is therefore no predicted nonlinear change to the frequency of the wave during the evolution. As discussed in Section \ref{sec: asymptotic solutions}, this model is not expected to capture chirping. The regime considered here with $\nu_{\text{eff}}^3/\omega_b^3 \gg \gamma/\omega_b$ is sufficiently collisional for all phase-space correlations to be erased on a timescale much faster than the mode amplitude is changing, and therefore no perturbations to the distribution function which would sustain chirping waves can persist in the system.   

\subsection{Comparison of the predicted phase with simulations}
\begin{figure} [h] 
    \begin{center}
    \includegraphics[width=0.75\textwidth]{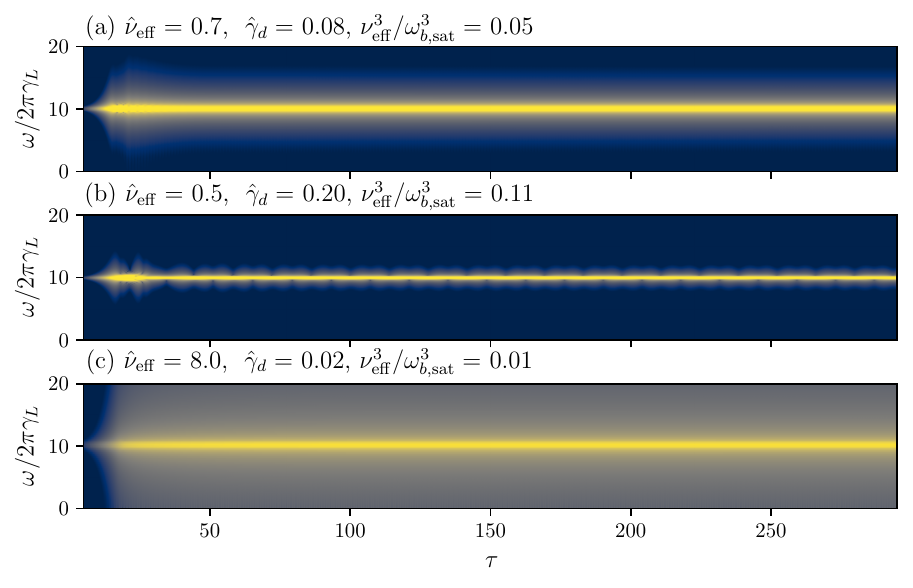}
    \end{center}
    \caption{Frequency spectrograms from BOT for the three example cases: (a) $\hat{\nu}_{\text{eff}} = 0.7$ and $\hat{\gamma}_d = 0.08$, (b)  $\hat{\nu}_{\text{eff}} = 0.5$ and $\hat{\gamma}_d = 0.2$, and (c) $\hat{\nu}_{\text{eff}} = 8.0$ and $\hat{\gamma}_d = 0.02$. The initial wave frequency is $\omega/2\pi\gamma_L=10$.}\label{fig: spectrograms}
\end{figure}

To verify the theoretical result that there should be no nonlinear change to the wave frequency, Fig. \ref{fig: spectrograms} shows  frequency spectrograms produced by BOT for the cases examined in the previous section. These also show no change to the wave frequency as the mode evolves in the small $\nu_{\text{eff}}^3/\omega_b^3$ regime. 

\section{Evolution of the distribution function} \label{sec: dist evolution}

\begin{figure}
    \begin{center}
    \includegraphics[width=1\textwidth]{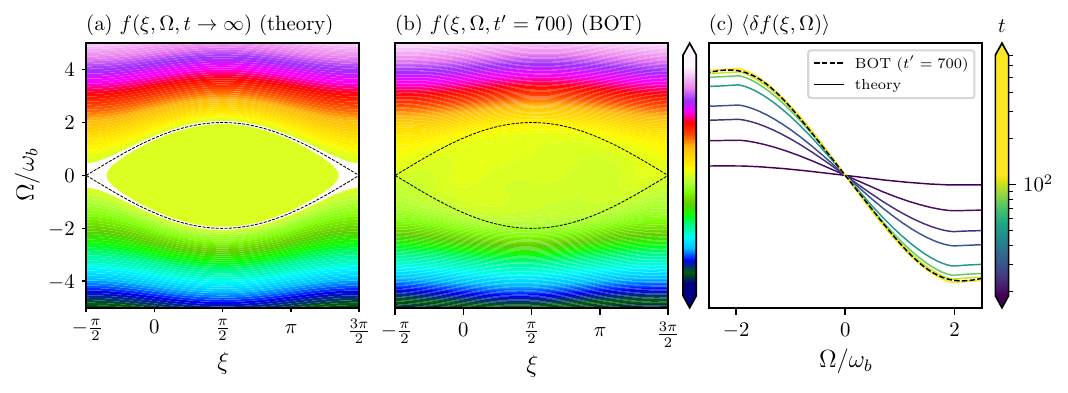}
    \end{center}
    \caption{The predicted distribution function in the original $(\xi, \Omega, t')$ coordinates compared with BOT for $\hat{\nu}_{\text{eff}} = 8.0$ and $\hat{\gamma}_d = 0.02$. (a) The distribution function $f \simeq f_0 + f_1$ predicted by Eqs. \ref{eq:f_0} and \ref{eq: f1} outside of the boundary layer. (b) The distribution function predicted by BOT. In both panels (a) and (b) the separatrix is plotted in dashed black. (c) The angle-averaged $ \langle \delta f(\xi,\Omega, t') = f_0(\xi,\Omega, t')+f_1(\xi,\Omega, t') - F_0(\Omega) \rangle$ computed from Eqs. \ref{eq:f_0} and \ref{eq: f1} at time steps ranging from $t' = t'_0 \approx 18$ (in navy) through $t' = 700$ (in yellow) plotted as solid lines. The saturated $\langle \delta f(z,\Omega)\rangle$ of the BOT simulation is shown in dashed black, and the initial deviation (at $t' = 0)$ from $F_0$ is shown in dotted red.  The width of the resonance is $\Delta \Omega /\omega_b = 4$.} \label{fig: delta_f}
 \end{figure}

Eq. \ref{eq: amplitude_solution} can also be used in the solution for $f$ to track the evolution of the distribution function as the amplitude evolves. Fig. \ref{fig: delta_f} shows the distribution function in the original $(\xi, \Omega, t')$ coordinates for the example case shown in panel (c) of Figs. \ref{fig: compare_BOT} and \ref{fig: spectrograms}.  Panel (a) shows the distribution function $f \simeq f_0 + f_1$ predicted by Eqs. \ref{eq:f_0} and \ref{eq: f1} outside of the separatrix boundary layer.  Panel (b) shows the distribution function predicted by BOT. Note that the result shown in panel (a) is effectively the distribution at $t' \rightarrow \infty$, whereas the BOT distribution is that at $t' = 700$, so some differences are expected, especially around the edges of the island where particles move very slowly. Panel (c) shows the angle averaged total change to the distribution function, defined by $\langle \delta f(\xi,\Omega, t) = f_0(\xi,\Omega, t')+f_1(\xi,\Omega, t) - F_0(\Omega)\rangle$.  This is computed numerically outside of the boundary layer\footnote{Since the boundary layer is narrow and we average over angle here, plotting the solution as if $\epsilon \rightarrow 0$ is likely to produce a nearly identical result to including it.} at several  times during the evolution, ranging from $t' = t'_0 \approx 18$ to $t' = 700$, when the mode has approximately reached saturation. The BOT $\delta f(\xi,\Omega)$ at $t' = 700$ is shown in dashed black, and initial perturbation is shown in dashed red.  

\section{BGK collision operator case} \label{sec: BGK}
\begin{figure}[H]
    \begin{center}
    \includegraphics[width=0.9\textwidth]{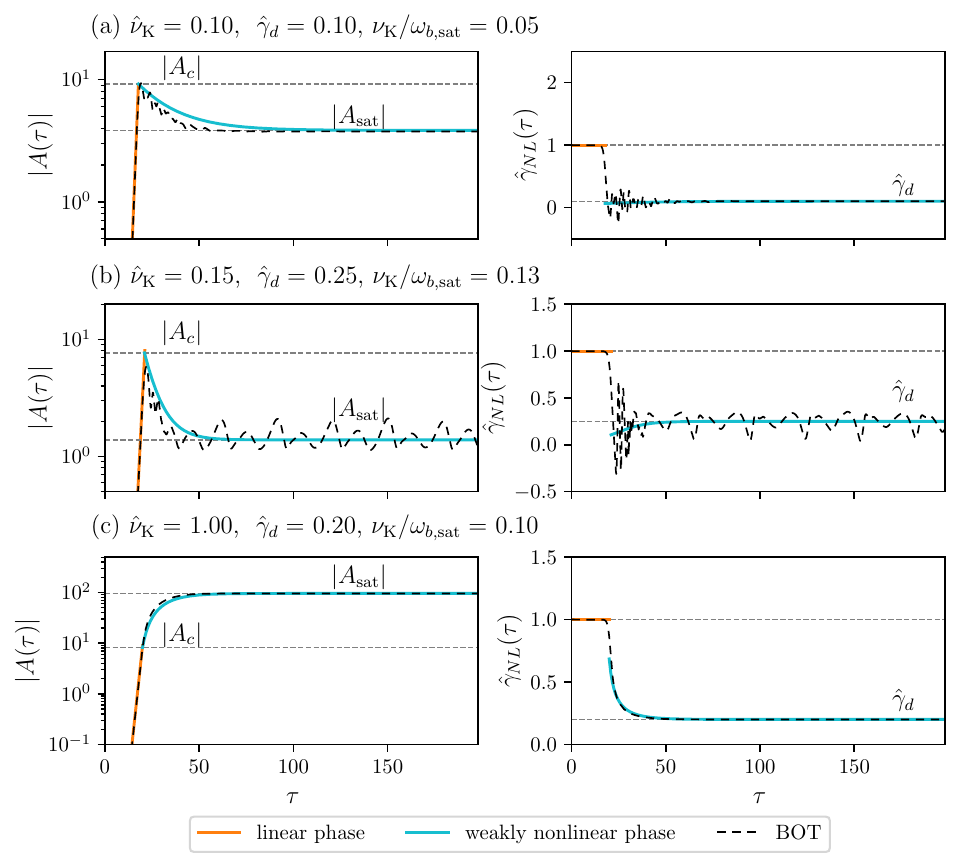}
    \end{center}  
    
    \caption{Comparison of theoretical predictions of Eqs. \ref{eq: amplitude_solution_krook} and \ref{eq: gammaNL_krook} for the amplitude and nonlinear growth rate with nonlinear kinetic simulations (using the BOT code) for three example cases: (a) $\hat{\nu}_{\text{K}} = 0.1$ and $\hat{\gamma}_d = 0.1$, (b) $\hat{\nu}_{\text{K}} = 1.0$ and $\hat{\gamma}_d = 0.2$, and (c) $
    \hat{\nu}_{\text{K}} = 0.5$ and $\hat{\gamma}_d = 0.5$. The left column shows the amplitude and the right column shows the nonlinear growth rate for each case.  The amplitude saturation level $|A_{\text{sat}}|=(1.9\hat{\nu}_{K}/\hat{\gamma}_{d})^2$ \cite{berk1990}, the collisionless saturation level $|A_c|=(3.2)^2$  \cite{fried1971}, and the final growth rate of $\hat{\gamma}_{NL}=\hat{\gamma}_d$ are shown in dotted black. Phase I (linear phase) is shown in orange and Phase II (weakly nonlinear phase) is shown in blue.}\label{fig: compare_BOT_krook}
\end{figure}

The saturation level for the same system in the case of a Krook or BGK annihilation operator, $C[\mathsf{F}] = -\nu_K \mathsf{F}$ \cite{bhatnagar1954}, and the source term $S(v) = \nu_K \mathsf{F}_0$, was found by Ref. \cite{berk1990} to be ${|A_\text{sat}|}^{1/2}=1.9\hat{\nu}_{K}/\hat{\gamma}_{d}$, where $\hat{\nu}_K$ is the normalized effective collision rate $\nu_K /\gamma_L$. In this case, the expansion parameter is $\nu_K/\omega_b \ll 1$. The mode amplitude in Phase II is then described by a slightly different Bernoulli ODE, 
\begin{equation} \label{eq: amplitude_krook}
    \frac{\mathrm{d}|A|}{\mathrm{d}\tau}+\frac{\gamma_{d}}{\gamma_L}|A|=1.9\hat{\nu}_{K}{|A|}^{1/2},
\end{equation}
which has the solution 
\begin{equation}\label{eq: amplitude_solution_krook}
    |A(\tau)|=\left[e^{-\hat{\gamma}_{d}(\tau-\tau_0)/2}\left({|A_{0}|}^{1/2}-{|A_\text{sat}|}^{1/2}\right)+{|A_\text{sat}|}^{1/2}\right]^2.
\end{equation} The nonlinear growth rate is given by 
\begin{equation} \label{eq: gammaNL_krook}
    \hat{\gamma}_{NL}(\tau) = \hat{\gamma}_d\left(\frac{|A_\text{sat}|}{|A(\tau)|}\right)^{1/2},  
\end{equation} where $|A(\tau)|$ is given by Eq. \ref{eq: amplitude_solution_krook}.
This result is compared with BOT simulations with the Krook operator in Figure \ref{fig: compare_BOT_krook}.  Panels (a) and (c) show cases with steady-state saturation levels where  Eq. \ref{eq: amplitude_solution_krook} agrees well with the simulations. Panel (b) shows a case with an oscillating solution, where the theoretical prediction captures only the average level.

Based on the results of Section \ref{sec: phase} and since we do not expect this model to capture chirping, we do not consider the phase equation for the BGK operator. BOT simulations show no changes to the frequency for the cases shown in Fig. \ref{fig: compare_BOT_krook}. 

\section{Conclusions}\label{sec: concl}
This study has established, for the first time, a simple analytic prediction of the nonlinear time evolution of the mode amplitude of a perturbatively driven kinetic instability far from its instability threshold. This work builds on the works of Ref. \cite{zakharov1963}, which calculated the rate of power exchange that occurs during the late stages of strongly driven interactions as a result of collisions, and Refs. \cite{berk1990,berk1990b, berk1990c,petviashvili1999}, which calculated the amplitude saturation levels for several collision operators. By using the separation of timescales between the linear and weakly nonlinear regimes and the strongly nonlinear transition between them, we have developed a theoretical prediction for the full time evolution of the mode amplitude in the far from threshold regime under a time-local, Markovian approximation. 

This prediction is derived under an assumption of sufficiently strong collisions or sufficiently slow nonlinear growth rate such that the kinetic equation can be expanded in the small parameter $\epsilon \equiv \nu_{\text{eff}}^3/\omega_b^3 \gg \gamma/\omega_b$.  Under these assumptions, the strongly driven regime is characterized first by a linear phase, sustained by energy from particles scattered into the resonance. When the mode amplitude becomes large enough for phase mixing to dominate over collisions, a fast strongly nonlinear transition occurs where the distribution function flattens around the resonance.  After this transition, the mode amplitude evolves in a weakly nonlinear regime to reach the final saturation level based on the balance of the kinetic drive and the wave dissipation. During this phase, the kinetic equation is shown to obey an advection-diffusion equation one dimensional phase space, which is a reduction in dimension from the original kinetic equation. The main results of this work, Eqs. \ref{eq: amplitude_solution} and \ref{eq: amplitude_solution_krook}, predict the mode amplitude during the weakly nonlinear phase, allowing for an analytic description of the full time evolution of the mode.  

This model accurately describes all features of the amplitude evolution for strongly driven cases in which the amplitude reaches a steady state saturation, as shown by comparison with nonlinear kinetic simulations using the BOT code. The theoretical prediction for the distribution function is also shown to match closely with these simulations. In cases where the amplitude reaches an oscillatory saturation level, the mode predicts the average level of those oscillations, as shown in Figs. \ref{fig: compare_BOT}(b) and \ref{fig: compare_BOT_krook}(b). The model also predicts no nonlinear changes to the wave frequency. Chirping will not appear in the regime considered by this model as the net mode growth rate cannot balance the collisional dynamics (i.e., $\nu_{\text{eff}}^3/\omega_b^3 \sim \gamma/\omega_b$). 

The strongly driven dynamics presented here are in contrast to the marginally unstable case, where the time-local regime leads to a purely diffusive transport equation with a resonance-broadened coefficient \cite{duarte2019}, instead of the advection-diffusion structure found here. In that case, the equilibrium distribution gradient is only slightly modified as the excitation remains marginal throughout its evolution due to $\gamma_d\sim\gamma_L$, and the mode amplitude satisfies a Landau-Stuart equation in the time-local limit \cite{duarte2019a}. All the nonlinear dynamics occur in a weakly nonlinear regime, allowing for the prediction of the mode amplitude with a single expression.  In the strongly driven case, analytic prediction is possible only by considering the linear and weakly nonlinear dynamics separately. These two phases of evolution are separated by a strongly nonlinear transition, and the solution for the amplitude in the weakly nonlinear regime is constructed using a perturbation theory approach on the distribution function which has already been strongly modified by the wave.  

\section{Applications} \label{sec: appl}
The reported results are directly applicable in many areas of plasma physics and beyond. Abrupt relaxation events in fusion plasmas can change the equilibrium and fast ion distributions on timescales shorter than the characteristic AE growth time, exciting AEs in the strongly driven regime. For example, in the JT-60U tokamak, strong excitation of ellipticity-induced AEs is seen following sawtooth crashes, attributed to the sudden reappearance of the $q=1$ surface resonant with an already formed fast ion distribution \cite{kramer2001}. Similar appearances of strongly driven AEs have been observed in DIII-D, TFTR, and JET \cite{fredrickson2000,fredrickson2000a,ruizruiz2025}, spanning several different AE types. These results are also relevant in modeling zonal flow excitation by drift wave turbulence, where the trapping of turbulent quasi-particles in a wave potential plays the role of the resonant distribution and provides kinetic drive for the wave \cite{mendonca2012}. Eqs. 6 and 15 of Ref. \cite{mendonca2012} for the kinetic equation and the power exchange between a zonal mode interacting with a turbulent bath are isomorphic to Eqs. \ref{eq: kinetic_Omega_xi} and \ref{eq: amplitude-Omega_xi} in this work. Additional plasma physics applications also include the modeling of lower hybrid current drive for an intense monochromatic field \cite{catto2025a}, and extending models of alpha particle loss in stellarators \cite{catto2025,catto2025b} to allow for time dependence and background damping.

As well as being broadly applicable in plasma physics, the results of this work are derived from governing equations which have isomorphisms in many other physical systems. In fluids, they may be used to describe instabilities in critical shear layers in viscous fluids in the strong nonlinear regime \cite{beland1978, maslowe1986}. For example, Eqs. 4.33 and 4.36 of Ref. \cite{goldstein1988} for the critical layer vorticity equation and the transverse jump condition that determines the instability amplitude are structurally identical to Eqs. \ref{eq: kinetic_Omega_xi} and \ref{eq: amplitude-Omega_xi} in this work. These results can therefore be used to explain the observed behavior of the amplitude evolution reported in simulations during the critical layer nonlinear vorticity roll-over phase \cite{goldstein1988}.

The results established here are also directly relevant in characterizing the evolution of resonant self-gravitating systems. Spiral instabilities in disk galaxies can behave similarly to an eigenmode in a plasma \cite{sellwood2020,hamilton2024}, exchanging energy with a sub-population of stars that satisfy a resonance condition. In this context, $\omega_{b}$ is the libration frequency of stars around the co-rotation resonance, the role of $\gamma_d$ is played by other subdominant (Lindblad) resonances, and $\nu_\text{eff}$ represents the diffusion of stellar orbits by the gaseous interstellar medium. Typical numbers for our Galaxy give $\Delta\equiv\nu_{\text{eff}}^3/\omega_{b}^3\approx0.2$ \cite{chiba2025a}, which is within the regime treated in this work. For example, Eqs. 30 and 53 of Ref. \cite{hamilton2023} for the transport equation for dark matter and the torque exerted by the resonant masses on the galactic bar are the same as our Eqs. \ref{eq: kinetic_Omega_xi} and \ref{eq: amplitude-Omega_xi}.  This analytic formulation can be used to predict and interpret the results of nonlinear simulations \cite{sellwood2020,chiba2025a} and to extend analytic results \cite{hamilton2024} to time-dependent scenarios with sources and sinks, for which the saturation level can be fundamentally different from the collisionless prediction.

\ack{We thank M. K. Lilley for developing and making the code BOT openly available; A. Bierwage and P. J. Catto for general comments on this manuscript; T. Barberis and C. Hamilton for pointing out the relevance of the strongly driven regime treated in this work to Alfv\'en eigenmode dynamics following sawtooth crashes tokamaks and to spiral instabilities in galaxies \cite{hamilton2024}, respectively; and E. D. Fredrickson and G. J. Kramer for clarifying discussions on the characteristic timescales for sawtooth relaxation and AE growth in TFTR, DIII-D and JT-60U.}

\funding{This manuscript is based upon work supported by the US Department of Energy, Office of Science, Office of Fusion Energy Sciences, and has been authored by Princeton University under Contract DE-AC02-09CH11466 with the US Department of Energy. The work was supported by the DOE Early Career Research Program, project \textit{Phase-Space Engineering of Supra-Thermal Particle Distribution for Optimizing Burning Plasma Scenarios}. The publisher, by accepting the article for publication, acknowledges that the United States Government retains a non-exclusive, paid-up, irrevocable, world-wide license to publish or reproduce the published form of this manuscript, or allow others to do so, for United States Government purposes.}




\appendix

\section{Derivation of the field amplitude equation} \label{appendix: amplitude equation derivation}
Considering an electrostatic wave and a single species in one dimension, the plasma current can approximated as $j = j_{\mathrm{cold}} +j_{\mathrm{res}}$. The contribution from the bulk plasma is $j_{\mathrm{cold}} = nq \mathrm{v}$, where $q$ is the charge, $n$ is the density and $\mathrm{v}$ is the fluid velocity. The contribution from the energetic resonant particles is $j_{\mathrm{res}} = q \int \mathrm{d}v \, vf(t,x,v)$, where $f(t,x,v)$ is the EP distribution function local to the resonance. Ampere's law with $B =0$ then reads
\begin{equation}\label{eq: ampere}
    \frac{\partial \mathcal{E}}{\partial t}  + 4\pi \left[nq \mathrm{v} +q \int \mathrm{d}v \, vf(t,x,v) \right] = 0,
\end{equation}
where $\mathcal{E}$ is the electric field. The fluid velocity $\mathrm{v}$ of the cold particles is described by the momentum equation 
\begin{equation} \label{eq: momentum}
    \frac{\mathrm{d} \mathrm{v}}{\mathrm{d} t} = \frac{q\mathcal{E}}{m} - 2\gamma_d \mathrm{v},
\end{equation}
where $m$ is the mass, and the last term models damping via collisional, Landau, or radiative processes at a constant rate $\gamma_d$. Combining Eqs. \ref{eq: ampere} and \ref{eq: momentum} to eliminate the fluid velocity $\mathrm{v}$, one obtains 
\begin{equation}\label{eq: amp der first}
    \frac{\partial^2 \mathcal{E}}{\partial t^2}+\omega_p^2 \mathcal{E} +2\gamma_d  \frac{\partial \mathcal{E}}{\partial t}=  - 4\pi q \frac{\partial}{\partial t}\int \mathrm{d}v \, vf(t,x,v)  - 8\pi q \gamma _d \int \mathrm{d}v \, vf(t,x,v),
\end{equation}
where $\omega_p= \sqrt{4\pi n q^2/m}$ is the plasma frequency. We assume an electric field of the form $\mathcal{E}(t,x) =  \frac{1}{2}[\tilde{\mathcal{E}}(t)\exp(ikx-i\omega t) + \mathrm{c.c.}]$, where $k$ is the wavenumber, $\omega$ is the real wave frequency, and $\tilde{\mathcal{E}}(t)$ is a slowly varying complex amplitude. It is assumed that $\partial/\partial t \sim \omega \gg \tilde{\mathcal{E}}'(t)$, and that $\partial/\partial t \sim \omega \gg \gamma_d$, such that the WKB approximation holds. The final term of Eq. \ref{eq: amp der first} is therefore much smaller than the others, and can be neglected.  Substituting in the form of the electric field to Eq. \ref{eq: amp der first}, and neglecting terms which contain second order derivatives of $\tilde{\mathcal{E}}(t)$, this becomes
\begin{equation} \label{eq: amp der second}
    \frac{1}{2}(\omega_p^2-\omega^2) \tilde{\mathcal{E}}(t) e^{i(kx-\omega t)} - i \omega \frac{\mathrm{d} \tilde{\mathcal{E}}}{\mathrm{d} t} e^{i(kx-\omega t)} - i \omega \gamma_d \tilde{\mathcal{E}}(t) e^{i(kx-\omega t)} +\mathrm{c.c.}\simeq i\omega 4\pi q \int \mathrm{d}v\, v f(t,x,v).
\end{equation}
Since the wave is electrostatic, $\omega_p=\omega$, so the first term vanishes.  Multiplying both sides by $\exp(-ikx)$ and averaging over one wavelength, we then find

\begin{equation}\label{eq: amp der third}
    \frac{\mathrm{d} \tilde{\mathcal{E}}}{\mathrm{d} t} +\gamma_d \tilde{\mathcal{E}} = - 2 q k \int_0^{2\pi/k} \mathrm{d}x \,e^{-i(kx-\omega t)} \int_{-\infty}^{\infty} \mathrm{d}v\, v f(t,x,v).
\end{equation}

\section{Bounce frequency}\label{appendix: bounce frequency}
For an electric field of the form $\mathcal{E}(t,x) = |\tilde{\mathcal{E}}(t)|\cos(kx-\omega t+\phi)$, the Hamiltonian describing the motion of a particle of mass $m$ and charge $q$ is given by
\begin{equation}
    H =  \frac{p^2}{2m}  - \frac{q|\tilde{\mathcal{E}}(t)|}{k}\sin(k x-\omega t+\phi),
\end{equation}
where $p = mv$. Assuming that $\phi$ changes slowly compared to the particle motion and defining $z = k x-\omega t+\phi$, the equation of motion of a particle is given by 
\begin{equation}
    \ddot{z} -\frac{q k|\tilde{\mathcal{E}}(t)|}{m} \cos z = 0.
\end{equation}
For particles with $q>0$, the O-point of the resonant island is located at $z = \pi/2$, and the $\cos z$ may be Taylor expanded around this point.  Deeply trapped particles therefore obey 
\begin{equation}
    \ddot{z} + \frac{q k|\tilde{\mathcal{E}}(t)|}{m}\left(z-\frac{\pi}{2}\right) \simeq 0,
\end{equation}
and their bounce frequency around the O-point is given by $\omega_b^2 \equiv qk|\tilde{\mathcal{E}}(t)|/m$.

\section{Transformation of coordinates in the kinetic equation} \label{appendix: coordinate transform}
The kinetic equation in the original space, velocity and time coordinates is given by

\begin{equation}\label{eqa: kinetic_Omega_xi}
    \frac{\partial f}{\partial t'}+\Omega\frac{\partial f}{\partial\xi}+|A(t')|\cos\xi\frac{\partial f}{\partial\Omega}=\hat{\nu}_{\text{eff}}^3\frac{\partial^2 f}{\partial \Omega^2}.
\end{equation}  Defining the new variables $\tau = t'$, $z = \xi+\phi(t')$, and particle energy $y = \Omega^2/(2|A(t')|)-\sin[\xi+\phi(t')]$, the derivatives in the kinetic equation become 
\begin{equation}
    \left( \frac{\partial f}{\partial t} \right)_{\xi, \Omega} = 
\left( \frac{\partial f}{\partial \tau} \right)_{z, y} + 
\phi'(\tau) \left( \frac{\partial f}{\partial z} \right)_{y, \tau} - 
\left( \frac{|A'(\tau)|}{|A(\tau)|}(y+\sin z) + \phi'(\tau) \cos z \right) \left( \frac{\partial f}{\partial y} \right)_{z, \tau}
\end{equation}

\begin{equation}
    \left( \frac{\partial f}{\partial \xi} \right)_{\Omega, t'} = 
\left( \frac{\partial f}{\partial z} \right)_{y, \tau} - 
\cos z \left( \frac{\partial f}{\partial y} \right)_{z, \tau}
\end{equation}

\begin{equation}
    \left( \frac{\partial f}{\partial \Omega} \right)_{\xi, t'} = 
\frac{\sqrt{2|A(\tau)|(y+\sin z)}}{|A(\tau)|} \left( \frac{\partial f}{\partial y} \right)_{z, \tau}
\end{equation}
where the upper and lower signs refer to $\Omega >0$ and $\Omega <0$, respectively. Using these expressions in Eq. \ref{eqa: kinetic_Omega_xi}, we have 
\begin{multline}\label{eqa: kinetic_y_z}
    \frac{\partial f^{\pm}}{\partial \tau} +\phi'(\tau)\frac{\partial f^{\pm}}{\partial z} \pm  \sqrt{2|A(\tau)|}\sqrt{y+\sin z} \frac{\partial f^{\pm}}{\partial z} \\-\left[\frac{|A'(\tau)|}{|A(\tau)|}(y+\sin z)+\phi'(\tau)\cos z \right]\frac{\partial f^{\pm}}{\partial y} = \frac{2\hat{\nu}_{\text{eff}}^3}{|A(\tau)|} \sqrt{y+\sin z}\frac{\partial}{\partial y}\sqrt{y+\sin z} \frac{\partial f^{\pm}}{\partial y},
\end{multline}

\section{First order constants of integration}\label{appendix: f1 integration constants}
The first order outer solutions for trapped and passing particles are given by 
\begin{equation} \label{ap: f1 trapped}
    f_1^{\pm(\mathrm{t})}(\tau,z,y)  = c_1^{\pm(\mathrm{t})}(\tau,y),
\end{equation} 
\begin{equation} \label{ap: f1 passing}
    f_1^{\pm(\mathrm{p})}(\tau, z,y)  = c_1^{\pm(\mathrm{p})}(\tau,y) + \frac{2\hat{\nu}_{\text{eff}}^3}{|A(\tau)|}\frac{\partial}{\partial y} \left(\frac{\int_{z_{1,2}}^{z} \mathrm{d}z'\sqrt{y+\sin z'}}{\int_{-\pi/2}^{3\pi/2} \mathrm{d}z \sqrt{y+\sin z}} \right)
\end{equation} 
where the upper and lower signs refer to $\Omega>0$ and $\Omega<0$ respectively. The constant of integration $c_1^{\pm(\mathrm{t})}(\tau,y)$ can be found by enforcing solvability conditions on the kinetic equation to next order, 
\begin{equation} \label{ap: kinetic_f2}
    \frac{\partial f_2^{\pm(\mathrm{t})}}{\partial z}  = \pm \sqrt{2}\frac{\hat{\nu}_{\text{eff}}^3}{|A(\tau)|^{3/2}}\frac{\partial}{\partial y}\sqrt{y+\sin z} \frac{\partial f_1^{\pm(\mathrm{t})}}{\partial y}.
\end{equation}
We now repeat the procedure used to find the form of $f_0^{\pm(\mathrm{t})}(\tau,y)$. For the trapped particles, Eq. \ref{ap: kinetic_f2} is integrated between the turning points, and again since there is no particle source at the O-point, we find $c_1^{\pm(\mathrm{t})}(\tau,y) = 0$.  

To find $c_1^{\pm(\mathrm{p})}(\tau,y)$, we use the condition that $f_1^{\pm(\mathrm{p})}(\tau,z,y)$ must be periodic with a zero-average in $z$.  Averaging Eq. \ref{ap: f1 passing} over $z$ yields 
\begin{equation} \label{ap: c1 passing}
     c_1^{\pm(\mathrm{p})}(\tau,y) =  \frac{\hat{\nu}_{\text{eff}}^3}{\pi |A(\tau)|}\frac{\partial}{\partial y}\underbrace{\left(\int_{-\pi/2}^{3\pi/2} \mathrm{d}z \frac{\int_{z_{1,2}}^{z} \mathrm{d}z'\sqrt{y+\sin z'}}{\int_{-\pi/2}^{3\pi/2} \mathrm{d}z \sqrt{y+\sin z}}\right)}_{\mathcal{H}^\pm}.
\end{equation} 
To calculate the function $\mathcal{H}^\pm$, it will be convenient to change variables to $s = \sin z$. The Jacobians of the $z$ integrals are then given by $\mathrm{d}z = \mathrm{d}s/\cos z = \pm \mathrm{d}s/\sqrt{1-s^2}$, where the sign is determined by the sign of $\cos z$. The integrals must be split at $\pi/2$, where the Jacobian changes sign, i.e,
\begin{equation}
    \mathcal{H}^\pm = \underbrace{\int_{-\pi/2}^{\pi/2}\mathrm{d}z \frac{\int_{z_{1,2}}^{z} \mathrm{d}z'\sqrt{y+\sin z'}}{\int_{-\pi/2}^{3\pi/2} \mathrm{d}z \sqrt{y+\sin z}} }_{\cos z \geq 0}  + \underbrace{\int_{\pi/2}^{3\pi/2}\mathrm{d}z \frac{\int_{z_{1,2}}^{z} \mathrm{d}z'\sqrt{y+\sin z'}}{\int_{-\pi/2}^{3\pi/2} \mathrm{d}z \sqrt{y+\sin z}} }_{\cos z \leq 0}.
\end{equation}
Making the change of variables yields 
\begin{eqnarray} \label{ap: H}
    \mathcal{H}^\pm &=&  \underbrace{\int_{-1}^{1}\frac{\mathrm{d}s}{\sqrt{1-s^2}} \frac{\mathcal{K}_1^\pm(s,y)}{\mathcal{K}_2(y)}}_{\cos z \geq 0}  - \underbrace{\int_{1}^{-1}\frac{\mathrm{d}s}{\sqrt{1-s^2}} \frac{\mathcal{K}_1^\pm(s,y)}{\mathcal{K}_2(y)}}_{\cos z \leq 0} \nonumber \\
    &=&  \int_{-1}^{1}\frac{\mathrm{d}s}{\sqrt{1-s^2}} \bigg(\underbrace{\frac{\mathcal{K}_1^\pm(s,y)}{\mathcal{K}_2(y)}}_{\cos z \geq 0} + \underbrace{\frac{\mathcal{K}_1^\pm(s,y)}{\mathcal{K}_2(y)}}_{\cos z \leq 0}\bigg)
\end{eqnarray}
where 
\begin{eqnarray} \label{ap: integraldefs}
    \mathcal{K}_1^+(s,y) &=&
    \begin{cases}\label{ap: integraldefs1}
        \displaystyle \int_{-1}^{s}\mathrm{d}s'\, \mathcal{G}(s',y), & \cos z \geq 0 \\
        \displaystyle \int_{-1}^{1}\mathrm{d}s\, \mathcal{G}(s,y)-\int_1^{s}\mathrm{d}s'\, \mathcal{G}(s',y), & \cos z \leq 0
    \end{cases}\\
    \mathcal{K}_1^-(s,y) &=&
    \begin{cases}\label{ap: integraldefs2}
        \displaystyle -\int_{-1}^{s}\mathrm{d}s'\, \mathcal{G}(s',y), & \cos z \leq 0 \\
        \displaystyle -\int_{-1}^{1}\mathrm{d}s\, \mathcal{G}(s,y) +\int_{1}^{s}\mathrm{d}s'\, \mathcal{G}(s',y), & \cos z \geq 0
    \end{cases}\\
    \mathcal{K}_2(y) &=& 2\int_{-1}^1\mathrm{d}s\, \mathcal{G}(s,y),\label{ap: integraldefs3}
\end{eqnarray}
and $\mathcal{G}(s,y) = \sqrt{y+s}/\sqrt{1-s^2}$. Substituting Eqs. \ref{ap: integraldefs1}-\ref{ap: integraldefs3} into Eq. \ref{ap: H}, yields 
\begin{equation}
    \mathcal{H}^\pm = \pm \pi.
\end{equation}
Therefore we find that 
\begin{equation} \label{ap: c1 passing final}
    c_1^{\pm(\mathrm{p})}(\tau,y) =  \frac{\hat{\nu}_{\text{eff}}^3}{\pi |A(\tau)|}\frac{\partial \mathcal{H}^\pm}{\partial y} = 0.
\end{equation} 

\section{Integration of the real amplitude equation}\label{appendix: real amplitude}
The real part of the mode amplitude satisfies (Eq. \ref{eq: amplitude_z_E_amp}) 
\begin{equation}\label{eq: real_amp start}
    \frac{\mathrm{d}|A|}{\mathrm{d}\tau}+\hat{\gamma}_{d}|A|=\frac{\sqrt{2|A|}}{\pi}\underbrace{\int_{-\pi/2+\phi}^{3\pi/2+\phi} \mathrm{d}z\int_{-\sin z}^{\infty}\mathrm{d}y\, \sqrt{y+\sin z} \frac{\partial}{\partial z}(f^+ +f^-),}_{\mathcal{I}}
\end{equation}
The integral may be split via the regions of the distribution function, i.e., $\mathcal{I} = \mathcal{I}^{(\mathrm{t})} + \mathcal{I}^{(\mathrm{iI})} + \mathcal{I}^{(\mathrm{iII})} + \mathcal{I}^{(\mathrm{p})}$, where 
\begin{eqnarray*}
    \mathcal{I}^{(\mathrm{t})} &=& \int_{-\pi/2+\phi}^{3\pi/2+\phi} \mathrm{d}z\int_{-\sin z}^{1-\epsilon^{1/2}}\mathrm{d}y\, \sqrt{y+\sin z} \frac{\partial}{\partial z}\left(f^{+(\mathrm{t})} +f^{-(\mathrm{t})}\right) \\ 
    \mathcal{I}^{(\mathrm{iI})} &=& \int_{-\pi/2+\epsilon^{1/4}}^{3\pi/2-\epsilon^{1/4}} \mathrm{d}z\int_{1-\epsilon^{1/2}}^{1+\epsilon^{1/2}}\mathrm{d}y\, \sqrt{y+\sin z} \frac{\partial}{\partial z}\left(f^{+(\mathrm{iI})} +f^{-(\mathrm{iI})}\right)\\
    \mathcal{I}^{(\mathrm{iII})} &=& \left(\int_{-\pi/2}^{-\pi/2+\epsilon^{1/4}} \mathrm{d}z +\int_{3\pi/2-\epsilon^{1/4}}^{3\pi/2} \mathrm{d}z\right) \int_{1-\epsilon^{1/2}}^{1+\epsilon^{1/2}}\mathrm{d}y\, \sqrt{y+\sin z} \frac{\partial}{\partial z}\left(f^{+(\mathrm{iII})} +f^{-(\mathrm{iII})}\right)\\
    \mathcal{I}^{(\mathrm{p})} &=& \int_{-\pi/2}^{3\pi/2} \mathrm{d}z\int_{1+\epsilon^{1/2}}^{\infty}\mathrm{d}y\, \sqrt{y+\sin z} \frac{\partial}{\partial z}\left(f^{+(\mathrm{p})} +f^{-(\mathrm{p})}\right).
\end{eqnarray*}
The contribution from particles in each region is calculated up to $\mathcal{O}(\epsilon)$ below.  

\begin{enumerate}
    \item \textit{Trapped particles:} The trapped particle distribution $f^{\pm(\mathrm{t})}$ has no dependence on $z$ to both zeroth and first order, so $\mathcal{I}^{(\mathrm{t})} \simeq \mathcal{I}_0^{(\mathrm{t})} +\mathcal{I}_1^{(\mathrm{t})} = 0$.
    \item \textit{Boundary layer particles in Region I:} Region I of the boundary layer contributes only at $\mathcal{O}(\epsilon)$, so $\mathcal{I}^{(\mathrm{iI})}\simeq \mathcal{I}_1^{(\mathrm{iI})}$. To leading order in the inner coordinate $\eta$ (recall $y = 1+\epsilon^{1/2} \eta$) the contribution from Region I is given by

    \begin{equation}
        \mathcal{I}^{(\mathrm{iI})} \simeq \epsilon^{1/2} \int_{-\pi/2 +\epsilon^{1/4}}^{3\pi/2-\epsilon^{1/4}} \mathrm{d}z\int_{1-\epsilon^{1/2}}^{1+\epsilon^{1/2}}\mathrm{d}\eta \, \sqrt{1+\sin z} \frac{\partial}{\partial z}\left(f_0^{+(\mathrm{iI})} +f_0^{-(\mathrm{iI})}\right).
    \end{equation}
    Substituting Eq. \ref{eq: reg1_f0}, this becomes
    \begin{eqnarray}
        \mathcal{I}^{(\mathrm{iI})} &=& \epsilon^{1/2} \sqrt{2} \int_{-\pi/2 +\epsilon^{1/4}}^{3\pi/2-\epsilon^{1/4}} \mathrm{d}z\int_{-1}^{1}\mathrm{d}\eta \, (1+\sin z)\left[\frac{\partial^2 f_0^{+(\mathrm{iI})}}{\partial \eta^2} - \frac{\partial^2 f_0^{-(\mathrm{iI})}}{\partial \eta^2}\right] \nonumber \\
        &=& \epsilon^{1/2} \sqrt{2} \int_{-\pi/2 +\epsilon^{1/4}}^{3\pi/2-\epsilon^{1/4}} \mathrm{d}z (1+\sin z)\left[\frac{\partial f_0^{+(\mathrm{iI})}}{\partial \eta} - \frac{\partial f_0^{-(\mathrm{iI})}}{\partial \eta}\right]_{-1}^{1}.
    \end{eqnarray}
    The distribution function must be continuous, so we may use the values of the outer solution derivatives at the edges of the boundary layer, which are independent of $z$.  Noting that $\partial_y f_0^{-(\mathrm{o})} = - \partial_y f_0^{+(\mathrm{o})}$, this gives
    \begin{eqnarray} \label{eq: Ii1}
        \mathcal{I}^{(\mathrm{iI})} 
        \simeq  4\pi \sqrt{2} \epsilon \frac{\partial f_0^{+(\mathrm{o})}}{\partial y}\bigg|_{1-\epsilon^{1/2}}^{1+\epsilon^{1/2}},
    \end{eqnarray}
    where terms smaller than $\mathcal{O}(\epsilon)$ have been neglected and where $\partial_y f_0^{+(\mathrm{o})}$ is given by Eq. \ref{eq:f_0 elliptic derivative}.
    \item \textit{Boundary layer particles in Region II:} In Region II of the boundary layer, the largest contribution is of $\mathcal{O}(\epsilon^{5/4})$ and therefore $\mathcal{I}^{(\mathrm{iII})} \simeq 0$. 
    \item \textit{Passing particles:} $f_0^{\pm(\mathrm{p})}$ is again independent of $z$, so $\mathcal{I}^{(\mathrm{p})} \simeq \mathcal{I}_1^{(\mathrm{p})} $. Substituting Eq. \ref{eq: kinetic_f1}, this yields 
    \begin{eqnarray}
        \mathcal{I}^{(\mathrm{p})} 
        &=& 2\sqrt{2}\epsilon\int_{-\pi/2}^{3\pi/2} \mathrm{d}z\int_{1+\epsilon^{1/2}}^{\infty}\mathrm{d}y\, \sqrt{y+\sin z}\frac{\partial}{\partial y}\left(\sqrt{y+\sin z} \frac{\partial f_0^{+(\mathrm{o})}}{\partial y}\right) \nonumber \\
        &=& 2\sqrt{2}\epsilon\int_{-\pi/2}^{3\pi/2} \mathrm{d}z\int_{1+\epsilon^{1/2}}^{\infty}\mathrm{d}y\,\left[(y+\sin z)\frac{\partial^2 f_0^{+(\mathrm{o})}}{\partial y^2} +\frac{1}{2}\frac{\partial f_0^{+(\mathrm{o})}}{\partial y} \right].
    \end{eqnarray}
    Evaluating the integrals over $z$ and integrating the first term by parts over $y$, this becomes 
    \begin{eqnarray}
        \mathcal{I}^{(\mathrm{p})} 
        &=&  4\pi\sqrt{2}\epsilon\int_{1+\epsilon^{1/2}}^{\infty}\mathrm{d}y\,\left(y\frac{\partial^2 f_0^{+(\mathrm{o})}}{\partial y^2} +\frac{1}{2}\frac{\partial f_0^{+(\mathrm{o})}}{\partial y} \right)\nonumber \\
        &=& 4\pi\sqrt{2}\epsilon \left(y\frac{\partial f_0^{+(\mathrm{o})}}{\partial y}\bigg|_{1+\epsilon^{1/2}}^{\infty} -\frac{1}{2}\int_{1+\epsilon^{1/2}}^{\infty}\mathrm{d}y\frac{\partial f_0^{+(\mathrm{o})}}{\partial y}\right).
    \end{eqnarray}
\end{enumerate}
Combining these contributions yields
\begin{eqnarray}
    \mathcal{I}
    &=& 4\pi \sqrt{2} \epsilon  \left(\frac{\partial f_0^{+(\mathrm{o})}}{\partial y}\bigg|_{1-\epsilon^{1/2}}^{1+\epsilon^{1/2}} +  y\frac{\partial f_0^{+(\mathrm{o})}}{\partial y}\bigg|_{1+\epsilon^{1/2}}^{\infty} -\frac{1}{2}\int_{1+\epsilon^{1/2}}^{\infty}\mathrm{d}y\frac{\partial f_0^{+(\mathrm{o})}}{\partial y}\right)
\end{eqnarray}
Finally, substituting Eq. \ref{eq:f_0 elliptic derivative}, and dropping terms of $\mathcal{O}(\epsilon^{3/2})$ originating at the boundaries gives
\begin{eqnarray}
    \mathcal{I}
    &\simeq& 8\pi\sqrt{|A|}  \epsilon \lim_{Y\rightarrow \infty}\left(\frac{Y}{\int_{-\pi/2}^{3\pi/2} \mathrm{d}z\, \sqrt{Y+\sin z}} -\frac{1}{2}\int_{1}^{Y}\frac{\mathrm{d}y}{\int_{-\pi/2}^{3\pi/2} \mathrm{d}z\, \sqrt{y+\sin z}}\right)  \nonumber \\
    &\approx& 3.90 \sqrt{|A|}  \epsilon.
\end{eqnarray}
Using $\epsilon \equiv \hat{\nu}_{\text{eff}}^3/|A|^{3/2}$, the real amplitude (Eq. \ref{eq: amplitude_z_E_amp})  therefore satisfies 
\begin{equation}\label{ap: amplitude_final}
    \frac{\mathrm{d}|A|}{\mathrm{d}\tau}+\hat{\gamma}_{d}|A|= 1.756 \frac{\hat{\nu}_{\text{eff}}^3}{|A|^{1/2}}.
\end{equation}

\section{Integration of the phase equation}\label{appendix: phase}
The wave phase satisfies (Eq. \ref{eq: amplitude_z_E_phase}),
\begin{multline} \label{eq: amplitude_z_E_phase_bp_app}
    \frac{\mathrm{d}\phi}{\mathrm{d}\tau}= \frac{\sqrt{2}}{\pi\sqrt{|A|}}\int_{-\pi/2+\phi}^{3\pi/2+\phi}\mathrm{d}z\bigg\{\lim_{y\rightarrow \infty}\left[\sin z\sqrt{y+\sin z}\left(f^{+(\mathrm{p})} +f^{-(\mathrm{p})}\right)\right] \\ - \int_{-\sin z}^{\infty} \mathrm{d} y\, \sin z\sqrt{y+\sin z} \frac{\partial }{\partial y}(f^+ +f^-)\bigg\}.  
\end{multline}
The first term is found to vanish: $f_1^{\pm(\mathrm{p})}(z,y\rightarrow \infty) \rightarrow 0$, and  the contribution from $f_0^{\pm(\mathrm{p})}$ is 
\begin{equation}
    \lim_{y\rightarrow \infty} \frac{2\omega}{\pi}\int_{-\pi/2}^{3\pi/2}\mathrm{d}z \sin z\sqrt{y+\sin z} = 0.
\end{equation}
Therefore, the phase is given by 
\begin{eqnarray} \label{eq: amplitude_z_E_phase_bp_app2}
    \frac{\mathrm{d}\phi}{\mathrm{d}\tau}= -\frac{\sqrt{2}}{\pi\sqrt{|A|}} \underbrace{\int_{-\pi/2+\phi}^{3\pi/2+\phi}\mathrm{d}z  \int_{-\sin z}^{\infty} \mathrm{d} y\, \sin z\sqrt{y+\sin z} \frac{\partial }{\partial y}(f^+ +f^-)}_{\mathcal{J}}.  
\end{eqnarray}
Again the integral will be split by region as $\mathcal{J} = \mathcal{J}^{(\mathrm{t})} + \mathcal{J}^{(\mathrm{iI})} + \mathcal{J}^{(\mathrm{iII})} + \mathcal{J}^{(\mathrm{p})}$, where 
\begin{eqnarray*}
    \mathcal{J}^{(\mathrm{t})} &=& \int_{-\pi/2+\phi}^{3\pi/2+\phi} \mathrm{d}z\int_{-\sin z}^{1-\epsilon^{1/2}}\mathrm{d} y\, \sin z\sqrt{y+\sin z} \frac{\partial }{\partial y}\left(f^{+(\mathrm{t})} +f^{-(\mathrm{t})}\right) \\ 
    \mathcal{J}^{(\mathrm{iI})} &=& \int_{-\pi/2+\epsilon^{1/4}}^{3\pi/2-\epsilon^{1/4}} \mathrm{d}z\int_{1-\epsilon^{1/2}}^{1+\epsilon^{1/2}}\mathrm{d} y\, \sin z\sqrt{y+\sin z} \frac{\partial }{\partial y}\left(f^{+(\mathrm{iI})} +f^{-(\mathrm{iI})}\right)\\
    \mathcal{J}^{(\mathrm{iII})} &=& \left(\int_{-\pi/2}^{-\pi/2+\epsilon^{1/4}} \mathrm{d}z +\int_{3\pi/2-\epsilon^{1/4}}^{3\pi/2} \mathrm{d}z\right) \int_{1-\epsilon^{1/2}}^{1+\epsilon^{1/2}}\mathrm{d} y\, \sin z\sqrt{y+\sin z} \frac{\partial }{\partial y}\left(f^{+(\mathrm{iII})} +f^{-(\mathrm{iII})}\right)\\
    \mathcal{J}^{(\mathrm{p})} &=& \int_{-\pi/2}^{3\pi/2} \mathrm{d}z\int_{1+\epsilon^{1/2}}^{\infty}\mathrm{d} y\, \sin z\sqrt{y+\sin z} \frac{\partial }{\partial y}\left(f^{+(\mathrm{p})} +f^{-(\mathrm{p})}\right).
\end{eqnarray*}
The contribution from particles in each region is calculated up to $\mathcal{O}(\epsilon)$ below.  
\begin{enumerate}
    \item \textit{Trapped particles:} Since both $\partial_y f_0^{\pm(\mathrm{t})}= 0$ and $\partial_y f_1^{\pm(\mathrm{t})}= 0$, $\mathcal{J}^{(\mathrm{t})} \simeq \mathcal{J}_0^{(\mathrm{t})} + \mathcal{J}_1^{(\mathrm{t})} =0$. 

    \item \textit{Boundary layer particles in Region I:} To zeroth order in the boundary layer coordinate $\eta$ (recall $y = 1+\epsilon^{1/2} \eta$), $\mathcal{J}^{(\mathrm{iI})}$ is given by
    \begin{eqnarray}
        \mathcal{J}^{(\mathrm{iI})} &\simeq& \int_{-\pi/2+\epsilon^{1/4}}^{3\pi/2-\epsilon^{1/4}} \mathrm{d}z\int_{-1}^{1}\mathrm{d} \eta\, \sin z\sqrt{1+\sin z} \frac{\partial }{\partial \eta}\left(f_0^{+(\mathrm{iI})} +f_0^{-(\mathrm{iI})}\right) \nonumber \\
        &=& \int_{-\pi/2+\epsilon^{1/4}}^{3\pi/2-\epsilon^{1/4}} \mathrm{d}z \sin z\sqrt{1+\sin z} \left(f_0^{+(\mathrm{iI})} +f_0^{-(\mathrm{iI})}\right)\bigg|_{-1}^1
    \end{eqnarray}
    The distribution is continuous, so we may again replace $f_0^{\pm(\mathrm{iI})}$ with $f_0^{\pm(\mathrm{o})}$ since it is evaluated only at the edges of the boundary layer.  This yields
    \begin{eqnarray}
        \mathcal{J}^{(\mathrm{iI})} &=& \int_{-\pi/2+\epsilon^{1/4}}^{3\pi/2-\epsilon^{1/4}} \mathrm{d}z\, \sin z\sqrt{1+\sin z} \left(f_0^{+(\mathrm{o})} +f_0^{-(\mathrm{o})}\right)\bigg|_{1-\epsilon^{1/2}}^{1+\epsilon^{1/2}} =0 
    \end{eqnarray}
    \item \textit{Boundary layer particles in Region II:} In Region II of the boundary layer, again the largest contribution is of $\mathcal{O}(\epsilon^{5/4})$ and therefore $\mathcal{J}^{(\mathrm{iII})} \simeq 0$. 
    \item \textit{Passing particles:} For the passing particles, we have $\mathcal{J}^{(\mathrm{p})} \simeq \mathcal{J}_0^{(\mathrm{p})}+\mathcal{J}_1^{(\mathrm{p})}$. The zeroth order is given by 
    \begin{equation}
        \mathcal{J}_0^{(\mathrm{p})} = \int_{-\pi/2}^{3\pi/2} \mathrm{d}z\int_{1+\epsilon^{1/2}}^{\infty}\mathrm{d} y\, \sin z\sqrt{y+\sin z} \frac{\partial }{\partial y}\left(f_0^{+(\mathrm{p})} +f_0^{-(\mathrm{p})}\right) = 0.
    \end{equation}
    The first order contribution is 
    \begin{equation}
        \mathcal{J}_1^{(\mathrm{p})} = \int_{-\pi/2}^{3\pi/2} \mathrm{d}z\int_{1+\epsilon^{1/2}}^{\infty}\mathrm{d} y\, \sin z\sqrt{y+\sin z} \frac{\partial }{\partial y}\left(f_1^{+(\mathrm{p})} +f_1^{-(\mathrm{p})}\right).
    \end{equation}
    It is convenient here to use the form of Eq. \ref{eq: f1 passing}, given by 
    \begin{equation} \label{eq: f1_app}
        f_1^{\pm(\mathrm{p})}(\tau, z,y)  = \frac{2\hat{\nu}_{\text{eff}}^3}{|A|}\frac{\partial}{\partial y} \left(\frac{\int_{z_{1,2}}^{z} \mathrm{d}z'\sqrt{y+\sin z'}}{\int_{-\pi/2}^{3\pi/2} \mathrm{d}z \sqrt{y+\sin z}} \right),
    \end{equation}
    where $z_1 = -\pi/2$ for particles with $\Omega>0$, and  $z_2 = 3 \pi/2$ for particles with $\Omega<0$. It will again be convenient to change variables to $s = \sin z$ and to use the form 
    \begin{equation}
        f_1^{\pm(\mathrm{p})}(s,y) = \frac{2\hat{\nu}_{\text{eff}}^3}{|A|} \frac{\partial }{\partial y}\left( \frac{\mathcal{K}_1^\pm(s,y)}{\mathcal{K}_2(y)}\right), 
    \end{equation}
    where $\mathcal{K}_1^\pm(s,y)$ and $\mathcal{K}_2(y)$ are given by Eqs. \ref{ap: integraldefs1}-\ref{ap: integraldefs3}. The Jacobian has two branches with opposite signs, so we must split the interval based on the sign of $\mathrm{d}z= \mathrm{d}s/\cos z = \mathrm{d}s/\pm\sqrt{1-s^2}$:
    \begin{equation}
        \mathcal{J}_1^{(\mathrm{p})} = \bigg(\underbrace{\int_{-\pi/2}^{\pi/2} \mathrm{d}z\int_{1+\epsilon^{1/2}}^{\infty}\mathrm{d} y }_{\cos z \geq 0} +  \underbrace{\int_{\pi/2}^{3\pi/2} \mathrm{d}z\int_{1+\epsilon^{1/2}}^{\infty}\mathrm{d} y }_{\cos z \leq 0}\bigg) \left[\sin z\sqrt{y+\sin z} \frac{\partial }{\partial y}\left(f_1^{+(\mathrm{p})} +f_1^{-(\mathrm{p})}\right) \right].
    \end{equation}
    This yields 
    \begin{multline}
        \mathcal{J}_1^{(\mathrm{p})} =\frac{2\hat{\nu}_{\text{eff}}^3}{|A|}\bigg[\underbrace{\int_{-1}^{1} \mathrm{d}s\int_{1+\epsilon^{1/2}}^{\infty} \frac{\mathrm{d} y\,s\sqrt{y+s}}{\sqrt{1-s^2}} \frac{\partial^2 }{\partial y^2}\left(\frac{\mathcal{K}_1^+(s,y)+\mathcal{K}_1^-(s,y)}{\mathcal{K}_2(y)}\right)}_{\cos z \geq 0} \\ -  \underbrace{\int_{1}^{-1} \mathrm{d}z\int_{1+\epsilon^{1/2}}^{\infty}\frac{\mathrm{d} y\,s\sqrt{y+s}}{\sqrt{1-s^2}} \frac{\partial^2 }{\partial y^2}\left(\frac{\mathcal{K}_1^+(s,y)+\mathcal{K}_1^-(s,y)}{\mathcal{K}_2(y)}\right)}_{\cos z \leq 0}\bigg].
    \end{multline}

    The integrals for the two regions can be combined 
    \begin{equation}
        \mathcal{J}_1^{(\mathrm{p})} =\frac{2\hat{\nu}_{\text{eff}}^3}{|A|}\int_{-1}^{1} \mathrm{d}s\int_{1+\epsilon^{1/2}}^{\infty} \frac{\mathrm{d} y\,s\sqrt{y+s}}{\sqrt{1-s^2}} \frac{\partial^2 }{\partial y^2}\bigg(\underbrace{\frac{\mathcal{K}_1^+(s,y)+\mathcal{K}_1^-(s,y)}{\mathcal{K}_2(y)}}_{\cos z \geq 0} + \underbrace{\frac{\mathcal{K}_1^+(s,y)+\mathcal{K}_1^-(s,y)}{\mathcal{K}_2(y)}}_{\cos z \leq 0}\bigg).
    \end{equation}
Substituting in the forms given by Eqs. \ref{ap: integraldefs1}-\ref{ap: integraldefs3}, we find that the quantity in the parenthesis vanishes.  Therefore, $\mathcal{J}_1^{(\mathrm{p})} = 0$.
\end{enumerate}
The phase equation to first order is therefore
\begin{equation}\label{ap: end_integration}
    \frac{\mathrm{d}\phi}{\mathrm{d}\tau}=0.
\end{equation}

\providecommand{\newblock}{}

\end{document}